\documentclass[fleqn,usenatbib]{mnras}

\usepackage{newtxtext,newtxmath}

\usepackage[T1]{fontenc}
\usepackage{ae,aecompl}
\usepackage{enumitem}
\usepackage{diagbox}

\newlist{inparaenum}{enumerate}{1}
\setlist[inparaenum]{nosep}
\setlist[inparaenum,1]{label=1\alph*)}

\newlist{num}{enumerate}{1}
\setlist[num]{nosep}
\setlist[num,1]{label=\arabic{numi})}

\newcommand{\ha}{{H$\alpha$}}
\newcommand{\gaia}{\textit{Gaia}}
\newcommand{\gIPHAS}{{\textit{Gaia}/IPHAS catalogue}}

\usepackage{graphicx}	
\usepackage{amsmath}	
\usepackage{mathtools}

\usepackage{array}
\usepackage{multirow}
\usepackage{lscape}
\usepackage{hyperref}

\title[Gaia/IPHAS catalogue of \ha-excess sources.]{Population-based identification of \ha-excess sources in the Gaia DR2 and IPHAS catalogues.}

\author[Fratta et al.]{M. Fratta,$^{1,2}$\thanks{E-mail: matteo.fratta@durham.ac.uk}
S. Scaringi,$^{1,2}$
J. E. Drew,$^{3}$
M. Mongui\'o,$^{4,5}$
C. Knigge,$^{7}$
\newauthor
T. J. Maccarone,$^{2}$
J. M. C. Court,$^{2}$
K. A. I\l{}kiewicz,$^{1,2}$
A. F. Pala,$^{6}$
P. Gandhi,$^{7}$
\newauthor
B. G{\"a}nsicke.$^{8}$
\\
$^{1}$Centre for Extragalactic Astronomy, Department of Physics, University of Durham, South Road, Durham, DH1 3LE, UK\\
$^{2}$Department of Physics and Astronomy, Texas Tech University, Lubbock, TX 79409-1051, USA\\
$^{3}$Dept of Physics and Astronomy, Faculty of Maths and Physical Sciences, University College London, Gower Street, London, WC1E 6BT, UK\\
$^{4}$Institut d'Estudis Espacials de Catalunya, Universitat de Barcelona (ICC-UB), Mart\'{i} i Franqu\`{e}s 1, E-08028, Spain\\
$^{5}$Universitat polit\`{e}cnica de Catalunya, Departament de F\'{i}sica, c/Esteve Terrades 5, 08860 Castelldefels, Spain\\
$^{6}$European Southern Observatory, Karl Schwarzschild Strasse 2, Garching bei Munchen, D-85748, Germany\\
$^{7}$School of Physics and Astronomy, University of Southampton, University Road, Southampton, SO17 1BJ, UK\\
$^{8}$Astronomy and Astrophysics Group, Department of Physics, University of Warwick, Gibbet Hill Road, Coventry, CV4 7AL, UK\\
}

\date{Accepted XXX. Received YYY; in original form ZZZ}

\pubyear{2021}

\usepackage{etoolbox}

\begin{document}
\label{firstpage}
\pagerange{\pageref{firstpage}--\pageref{lastpage}}
\maketitle

\begin{abstract}
\large
We present a catalogue of point-like \ha-excess sources in the Northern Galactic Plane. Our catalogue is created using a new technique that leverages astrometric and photomeric information from \gaia\ to select \ha-bright outliers in the INT Photometric \ha\ Survey of the Northern Galactic Plane (IPHAS), across the colour-absolute magnitude diagram.
To mitigate the selection biases due to stellar population mixing and to extinction, the investigated objects are first partitioned with respect to their positions in the \gaia\ colour-absolute magnitude space, and in the Galactic coordinates space, respectively. The selection is then performed on both partition types independently. Two significance parameters are assigned to each target, one for each partition type. These represent a quantitative degree of confidence that the given source is a reliable \ha-excess candidate, with reference to the other objects in the corresponding partition. Our catalogue provides two flags for each source, both indicating the significance level of the \ha-excess. By analysing their intensity in the \ha\ narrow band, 28,496 objects out of 7,474,835 are identified as \ha-excess candidates with a significance higher than 3. The \textit{completeness} fraction of the \ha\ outliers selection is between $3\%$ and $5\%$. The suggested $5\sigma$ conservative cut yields a \textit{purity} fraction of $81.9\%$.
\end{abstract}




\section{Introduction}
\label{sec: introduction}
\large
\ha\ emission can be observed from both extended sources, such as nebulosities associated with either star-forming regions and/or stellar remnants, and from point-like sources, with no associated extended emission.
These latter objects can fall into different source-types and can span various evolutionary stages of stellar populations. The many classes of \ha\ emitting point-like sources include (but are not limited to) a wide range of young stellar-objects (YSOs), classical Be stars, compact planetary nebulae, luminous blue variables (LBVs), hypergiants, Wolf-Rayet stars, and rapidly rotating stars. Furthermore, many interacting binary systems exhibit \ha\ in emission due to accretion (e.g. Cataclysmic Variables; CVs, Symbiotic Stars; SySt, or binary systems in which the accreting compact object is a Black Hole or a Neutron Star). The \ha\ emitting population is heterogeneous and challenging to identify. Because of this, samples of these objects are plagued by selection biases, which in turn prevent stellar evolution models from being adequately tested.

Large, wide-field, high angular resolution \ha\ imaging surveys provide the basis to discover and characterise \ha-excess sources. Among the previous surveys targeting the ionised diffuse interstellar medium (ISM) that have aimed to increase the sample of known \ha\ sources we can include, for instance, the \ha\ observations of the Large and Small Magellanic Clouds \citep{davies}. However, this survey only observed small patches of the sky, and the limiting magnitude was quite stringent. On the other hand, the Virginia Tech \ha\ and [S\textsubscript{II}] Imaging Survey of the Northern Sky (VTSS, \citealt{dennison}) and the Southern \ha\ Sky Survey Atlas (SHASSA, \citealt{gaustad}), covered wider areas of sky, but they suffered from relatively poor angular resolution. Among the imaging surveys that focused on point sources, \cite{kw99} obtained a list of $\sim$4000 point-like \ha\ emitters located in the northern Galactic plane ($|b|\leq10^\circ$). \cite{shs}, with their Anglo-Australian Observatory/UK Schmidt Telescope (AAO/UKST) SuperCOSMOS \ha\ Survey (SHS), inspected an area of $\sim4000$\,deg$^2$ in the Southern Milky Way, plus an additional $\sim700$\,deg$^2$ area around the Magellanic Clouds.

The Isaac Newton Telescope (INT) Photometric \ha\ Survey of the Northern Galactic Plane (IPHAS, \citealt{drew}) provides photometry with the 2 broad-band $r$ and $i$ filters, as well as with the narrow-band \ha\ filter (see also \citealt{drew} and \citealt{irwin}). \cite{witham} used the IPHAS pre-publication photometric measurements (without a uniform calibration) to identify candidate \ha\ emission line sources.
Their method is based on producing two-colours diagrams (TCDs) for each IPHAS field, using $r-$\ha\ and $r-i$, respectively, as vertical and horizontal axes. Each Wide Field Camera (WFC) pointing covers an area of 0.22\,deg$^2$ in the sky. \ha\ line excess source candidates are then selected by iteratively fitting the stellar locus and retaining positive outliers in $r-$\ha. This procedure is performed within pre-defined magnitude ranges to try and mitigate the effect of extinction, which can become substantial when looking through the Galactic plane \citep{sale}. Using their conservative method, \citealt{witham} identified in total 4,853 \ha\ emitting candidates.
Only a small fraction of these candidates could be confirmed through a comparison with previously developed narrow-line emitters catalogues. A spectroscopic follow-up (presented in \citealp{raddi}) was then performed on 370 outliers with $r<18$, and 97\% of them did show \ha\ emission lines.

More recently, \cite{monguio} developed the IGAPS (INT Galactic Plane Survey) catalogue, that includes $\sim295$\,millions objects. Of these, 53,234,833 (18\%) unblended sources with $r < 19.5$\,mag were tested for \ha-excess. IGAPS consists of a cross-match between IPHAS and UVEX (the UV-Excess survey of the Northern Galactic Plane, \citealt{uvex}). With the use of the 2.5\,m INT, the latter survey provides photometric measurements for the sources included in a 10\textdegree\ $\times$ 185\textdegree\ sky area, centred on the Galactic Equator. More specifically, it provides $U$, $g$ and $r$ intensities, with a limiting magnitude of 21-22\,mag. The $g$, $r$ and $i$ magnitudes in IGAPS were calibrated with reference to the ``Pan-STARRS photometric reference ladder" \citep{magnier}, while the \ha\ narrow-band calibration was based on the methods described in \cite{glazebrook}.

\smallskip

In the context of IPHAS, the main metric for \ha-excess is $r-$\ha. However, this colour-index is not quite constant for stars without emission lines, but varies as a function of the spectral type.
Without first confining distinct populations, the measured \ha\ excess of a star in the IPHAS TCD cannot have a consistent relation with the net emission equivalent width, and candidates can remain lost in the main stellar locus. For this reason, population-based \ha-excess selections generally produce more complete results. An example of such study is presented in \cite{mohr}: these authors performed their selection of \ha-excess candidates on a set of previously identified O and early B stars, across the Carina Arm. Their goal was an assessment of the relative frequency of the Classical Be (CBe) phenomenon in the VST Photometric \ha\ Survey of the Southern Galactic plane and Bulge (VPHAS+, \citealp{vphas}) field of view.

Without any knowledge of the distances, and using only IPHAS measurements, degeneracies may exist in associating a particular object to a specific stellar population. Because of this, the emission line candidate lists of \cite{witham} and \cite{monguio} are necessarily conservative and incomplete. In our work, \ha\ line excess candidates are identified from IPHAS survey by using two independent and complementary methods: a) selecting \ha-excess sources relative to nearby sources in the calibrated \gaia\ colour-absolute magnitude diagram (CAMD), and b) selecting \ha-excess sources relative to groups of objects that occupy nearby positions in the sky.
It is relevant to stress the fact that the objects that are labelled as \ha\ line excess candidates in this study are not necessarily \ha\ emitters; the only conclusion that can be reached through this selection process is that their \ha\ intensity is higher than that associated with objects they are compared to.

The input catalogue used in this work to identify \ha-excess sources, is that of \citealp{Scaringi} (hereafter \gIPHAS), which is the result of a positional sub-arcsecond cross-match between the sources in the \gaia\ and IPHAS DR2 fields of view. When performing the cross-match, \cite{Scaringi} took into account the proper motions provided by \gaia\, in order to rewind the positions of the objects back to the IPHAS DR2 observation epoch. This catalogue contains a list of approximately 8\,million sources, all found in the Northern Galactic plane.

\smallskip

In section \ref{sec: input}, a more detailed description of the input catalogue is provided. Section \ref{sec: selection} consists of an explanation of our selection process. The results obtained by our algorithm are presented in section \ref{sec: results}. In section \ref{sec: discussion} these results are discussed. Section \ref{sec: follow-ups} presents two possible science cases. In section \ref{sec: conclusion} we draw our conclusions.

\section{The input catalogue}
\label{sec: input}
The targets in the \gIPHAS\ occupy an area of the sky included between $|b|\leq 5$\textdegree\ and 29\textdegree $\leq l \leq$ 215\textdegree, and are mostly found within a distance radius of $\sim1.5$\,kpc from us. These distances are calculated directly as the inverse of \gaia\ parallax measurements, with the caveat that they satisfy the $parallax\_over\_error > 5$ criterion (\citealp{Scaringi}; median parallax uncertainties as well as the systematic parallax offset are discussed in \citealp{lindgren}).
The choice of inferring the distances via parallax inversion is justified by \cite{Scaringi} with the introduction of two parameters that quantify the goodness of \gaia\ astrometric fit and the false-positive rate: f\textsubscript{c} and f\textsubscript{FP}, respectively. To compute f\textsubscript{c}, they binned the targets according to their \gaia\ G band magnitudes; f\textsubscript{c} corresponds to the percentile assigned to each object in the bin, with respect to the $\chi^2$ of the astrometric fit. On the other hand, f\textsubscript{FP} reflects the presence of spurious negative parallaxes in \gaia\ measurements, due to poor astrometric fits. To obtain f\textsubscript{FP}, \cite{Scaringi} first produced a mirror sample of their catalogue, including only objects with negative parallaxes, with ``$parallax\_over\_error < - 5$". They thus binned the objects in the catalogue (including the mirror sample) with respect to their G band measurements, and further with respect to the $\chi^2$ of their astrometric fit. They thus define f\textsubscript{FP} as the fraction of objects from the mirror sample (false positives) in each bin.

To obtain the absolute magnitude for the \gaia\ G band (M\textsubscript{G}), \cite{Scaringi} used the distances calculated with the parallax-inversion method. Despite the precautions taken, this approximation contributes to the uncertainties on M\textsubscript{G}. However, the effects on M\textsubscript{G} introduced by the use of parallax-inversion method instead of probabilistic methods \citep{dist} to obtain the distances are generally negligible. In fact, 97.2\% of the objects in our meta-catalogue fall in the $|\delta_{M_G}|\leq 0.1$\,mag range, $\delta_{M_G}$ being the difference between the G-band absolute magnitudes obtained with the parallax-inversion defined distances and with probabilistically defined distances\footnote{The absolute value of the maximum difference is $|\delta_{M_G, max}|=0.36$\,mag.}.

Besides the errors on \gaia\ photometric measurements and the effects connected to the parallax-inversion defined distances, the location of the sources in the CAMD is also affected by the different extinctions that alter their colours. All these uncertainties may be the causes of stellar population mixing. Our approach to overcome this obstacle is presented in the last paragraph of Sec.~\ref{sec: CMD-partitions}.

\subsection{Additional Data quality constraints}
\label{sec:cuts}
This work focuses on the subset of targets from the \gIPHAS\ that pass strict quality criteria, in order to minimise the inclusion of spurious cross-matches. Some quality control selection cuts have already been applied during the compilation of the \gIPHAS, which are mostly aimed at retaining only those sources with good \gaia\ parallax measurements and good IPHAS photometry. Additional cuts are applied here, in order to:
\begin{enumerate}
    \item remove sources with low-quality astrometric fits and/or high false-positive probabilities (see Sec.~\ref{sec: input});
    \item remove targets close to t he saturation limit of IPHAS;
    \item only retain targets for which we have a valid measurement in each band of interest ($r$, $i$, \ha, M\textsubscript{G}, G\textsubscript{BP} and G\textsubscript{RP}).
\end{enumerate}
The following cuts are thus applied:
\begin{enumerate}
    \item retain sources that satisfy both
    $f_c<0.98$ and $f_{FP}\leq0.02$ (as suggested in \citealp{Scaringi});
    \item retain sources with
    $r\geq13$\,mag, $i\geq12$\,mag and \ha$\geq12.5$\,mag;
    \item retain only sources with measurements in all $r$, $i$, \ha, M\textsubscript{G}, G\textsubscript{BP} and G\textsubscript{RP} bands.
\end{enumerate}

\begin{figure}
	\includegraphics[width=5\columnwidth/6, trim= 0mm -5mm 0mm 0mm]{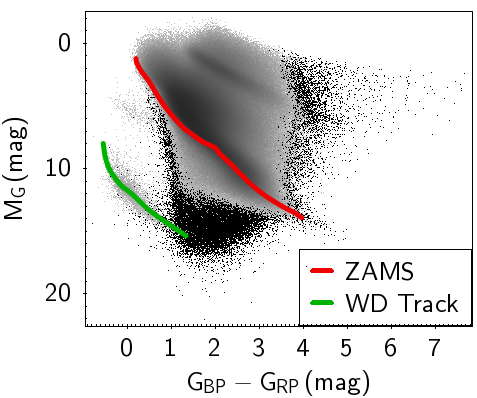}
	\includegraphics[width=5\columnwidth/6, trim= 12mm 3mm 4mm 0mm]{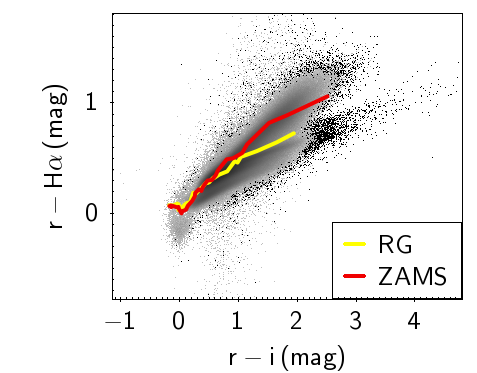}

    \caption{\normalsize Positions in the \gaia\ M\textsubscript{G} vs. G\textsubscript{BP}-G\textsubscript{RP} CAMD (top panel) and in the IPHAS $r-$\ha\ vs. $r-i$ TCD (bottom panel) of the sources in the \gIPHAS\ \protect\citep{Scaringi}. The grey dots represent the objects that satisfy the quality constraints described in Sec.~\ref{sec:cuts}, while the targets that do not pass this first selection are displayed with the black dots. The red and the green lines in the top panel represent respectively the synthetic Zero Age Main Sequence (ZAMS) track \protect\citep{Bressan} and the synthetic white dwarfs track \protect\citep{Carrasco}. The red line and the yellow line in the bottom panel (both taken from \protect\citealp{drew}) depict respectively the synthetic ZAMS track for zero reddening and the synthetic Red Giant (RG) track, in this parameter space.}
    
    \label{fig:CMD_CCD}
\end{figure}

These cuts yield 7,474,835 sources out of the original 7,927,224. Fig.~\ref{fig:CMD_CCD} shows the \textit{Gaia} colour-absolute magnitude diagram (i.e. the \gaia\ Hertzsprung-Russell diagram; HRD) and IPHAS two-colours diagram with the targets that pass the additional quality cuts.

\section{Selecting \texorpdfstring{\ha}{}-excess source candidates}
\label{sec: selection}

The aim of this work is to identify \ha-excess candidates in a vast sample of objects. This task is achieved by selecting ``positive outliers'' in the $r-$\ha\ vs. $r-i$ two-colours space. To mitigate the selection biases due to stellar population mixing and to Galactic extinction, the sources in the master-catalogue are first partitioned with respect to their positions in the \gaia\ CAMD and in the Galactic coordinates space, respectively. The two $r-$\ha\ outliers selections, performed on the CAMD-based and on the coordinates-based (or also ``positional"-based) partitions, are independent and complementary. The selection strategy performed on the coordinates-based partitions hinges on the one applied by \citealp{witham}. We point out that our CAMD-based selection can still be improved, since some populations may overlap in the colour-absolute magnitude space.

It is worth pointing out that other techniques using more novel machine learning approaches could be employed for the selection of \ha-excess sources. Our choice of a more rational approach is based on the relative simplicity of the algorithm, which allows to locate exactly in which partition a specific source has been selected from. Furthermore the approach used here allows us to examine and understand the underlying population used to infer the \ha-excess significance values.

The separation of the sources in the two parameter spaces is described in section \ref{sec: partition}, whilst the proper selection of \ha-excess sources is discussed in section \ref{sec:detrending and selection}.

\subsection{Partitioning algorithms}
\label{sec: partition}
\subsubsection{CAMD-based partitions}
\label{sec: CMD-partitions}

Using the calibrated \gaia\ CAMD shown in the top panel in Fig.~\ref{fig:CMD_CCD}, subsets (i.e. the partitions) are defined such that they a) contain a large enough number of sources (500) to be able to statistically identify outliers and b) are small enough in colour-absolute magnitude space to make the underlying source population as homogeneous as possible. To balance these two requisites, an iterative method is applied.

\begin{figure}
    \includegraphics[width=8\columnwidth/9, trim= 0mm 0mm 8mm 0mm]{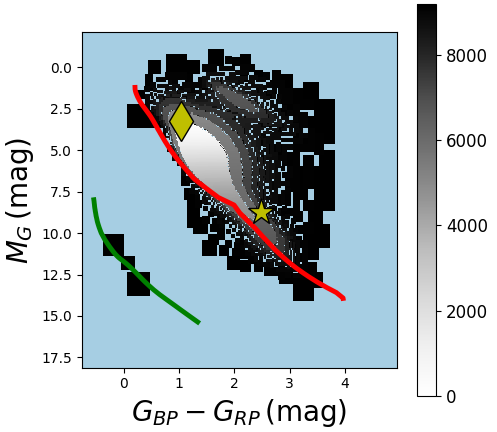}
    \caption{\normalsize Map of the partitions in the \gaia\ CAMD. The colour code refers to the order in which the partitions were created (no partitions are assigned to the light blue area). The area covered by the single partition increases where the density of sources decreases. The red line represents the synthetic ZAMS track \protect\citep{Bressan}, while the green line depicts synthetic white dwarfs track \protect\citep{Carrasco}.
    The yellow star points to partition 7331, while the diamond refers to partition 0, which are discussed in section \ref{sec:detrending and selection}.}
    \label{fig:partitions}
\end{figure}

First, a fine grid of $840 \times 840$ equally spaced ``elemental" cells is generated, covering the whole CAMD (each elemental cell with dimensions $l\textsubscript{x}$ $\sim0.007$\,mag and $l\textsubscript{y}$ $\sim0.024$\,mag, respectively). No elemental cells contain enough objects to be considered a partition. The side lengths of the grid cells are then increased to the next integer divisor of 840, in units of l\textsubscript{x} and l\textsubscript{y}, respectively.
The second iteration produces $420 \times 420$ cells, 4,852 of which satisfy the criteria to become partitions (these belong to the densest regions of the CAMD). These partitions are labelled according to the order by which they are generated during the current iteration (left to right, top to bottom), from 0 to 4,851. 
The iterations carry on for all the integer divisors of 840, between 2 and 60\footnote{Caveat: each partition must not be completely surrounded by another one; furthermore, all the elemental cells within each partition must be contiguous.}. 
At the end of the iterative procedure, 204,459 objects are still left without a partition assignment. These ``leftovers" are assigned to the closest partition. 
9,181 $CAMD-partitions$ result from this process, with a maximum density of 1,514 sources per partition. The map of the resulting partitions in the CAMD is shown in Fig.~\ref{fig:partitions}

To account for the uncertainty on the positions of the objects in the CAMD, this partitioning process is repeated upon change of the side lengths of the elemental cells. 
As an example, a 20\% increase of the side lengths of the elemental cells produces a 0.8\% variation in the number of selected outliers, meaning that our selection is independent (to a reasonable extent) on the size of the elemental cells.

\subsubsection{Coordinates-based partitions}
\label{sec: positional-partitions}

For this different partitioning algorithm, an evenly spaced grid in the $b$ vs. $l$ space is created. The size of each cell is 1.205\textdegree\ $\times$ 1.004\textdegree, i.e. about five times bigger than the ``cell size" used by \citealp{witham} (which performed their selection on an IPHAS field-by-field basis), and is chosen so that all the cells are either empty, or contain at least 500 objects. This procedure results in 1,674 $positional-partitions$, with a maximum density of 12,604, and a minimum density of 546 objects per partition.

\subsection{Detrending and identification of outliers.}
\label{sec:detrending and selection}
The $r-$\ha\ vs. $r-i$ TCDs are used to identify \ha\ line excess sources from every partition. First the main stellar population locus is found in each partition by iteratively fitting a line to the data, and applying Chauvenet's criterion. The latter consists of calculating a threshold\footnote{Chauvenet's threshold depends on the root mean square of the distribution and on the number of objects that constitute such distribution.} beyond which only outliers are expected to be found. The outliers are removed from the data at the end of each iteration. In theory, in order to tackle the population-mixing issue, the fit should be forced to the upper branch in the TCD (as done by \citealp{witham}) by removing only the negative outliers. However, for most of the partitions the resulting best-fit line does not deviate sensibly from the model obtained by the direct application of the unmodified Chauvenet's criterion.

\begin{figure*}
	\includegraphics[width=\textwidth/2, trim= 0mm 0mm 0mm 0mm]{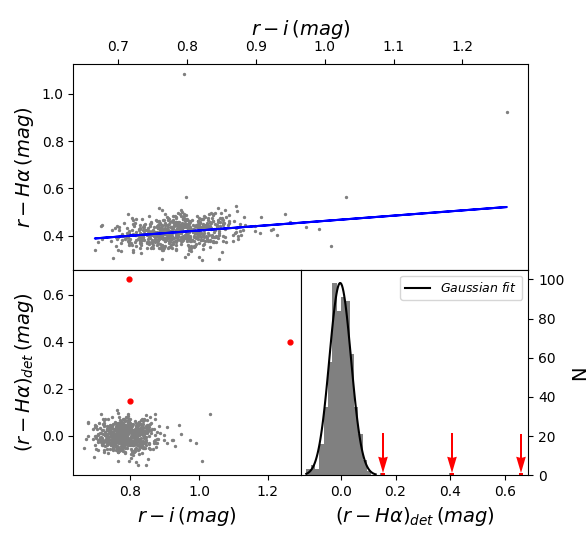}
    \caption{\normalsize Graphical depiction of the detrending and outliers selection processes performed on CAMD-partition 7331. The top panel shows the corresponding non-detrended IPHAS TCD, in which only one linear trend is clearly visible. The blue line depicts the best-fit linear model. It was subtracted from the data points to obtain the detrended $r$-\ha\ parameter (bottom-left panel). The red dots represent the positive outliers of the distribution. The bottom-right panel shows the detrended $r$-\ha\ distribution of this partition. The Gaussian behaviour of the underlying population is well described by the best-fit model (the black solid line). The red arrows point to the three outliers of the distribution.
    The position of the CAMD-partition 7331 in the \gaia\ CAMD is shown in Fig.~\ref{fig:partitions}.}
    \label{fig:composita7331}
\end{figure*}

Once the stellar locus has been located, it is used as a baseline to identify the outliers: each TCD is detrended by subtracting the corresponding linear model from the data. A second iterative application of Chauvenet's criterion on the detrended TCD enables us to isolate the outliers, and hence to calculate the standard deviation (\textit{rms}) of the remaining sources. The objects that satisfy the following relation are selected as \ha-excess candidates, from either the CAMD-based and/or the positional-based partitions:
\begin{equation}
    \sigma=\frac{y}{\sqrt{(\delta y)^2 + (m_{fit}\cdot \delta x)^2 + rms^2}}\geq3.
	\label{eq:selection}
\end{equation}
Here, $y$ corresponds to the $(r-$\ha)\textsubscript{detrended} intensity, $\delta y$ is the instrumental uncertainty on this value, $\delta x$ is the instrumental error on the $r-i$ intensity, and $m_{fit}$ is the slope of the best-fit line. Thus defined, $\sigma$, or \textit{significance}, represents the confidence that each source is an outlier of the corresponding distribution. Since the partitioning process is implemented in two different parameter spaces, two significances are assigned to each source: $CAMD-significance$ ($\sigma_{CAMD}$) and $POS-significance$ ($\sigma_{POS}$). Objects that satisfy relation \ref{eq:selection}, either from the CAMD-based and/or from the positional-based selection, will henceforth be referred to as ``$3\sigma$ outliers". Fig.~\ref{fig:composita7331} provides a graphical depiction of the detrending (top panel) and selection (bottom-left panel) processes relative to the CAMD-partition 7331, as an example of a well-behaved partition.

\begin{figure*}
	\includegraphics[width=\textwidth/2, trim= 0mm 0mm 0mm 0mm]{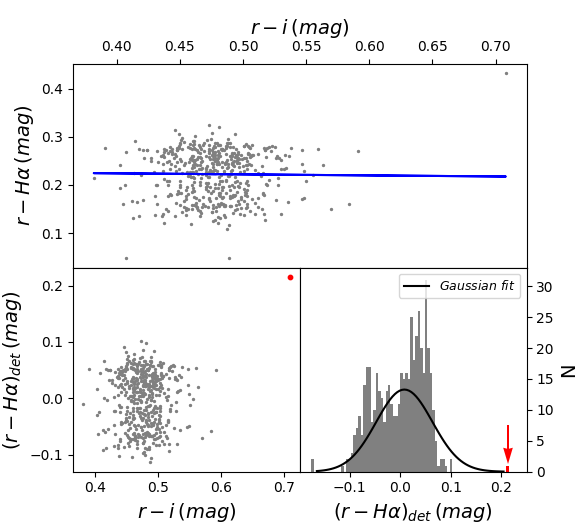}
    \caption{\normalsize Graphical depiction of the detrending and outliers selection processes performed on CAMD-partition 0. As it stands out clearly, the Gaussian model is not a good fit to the underlying population.
    The position of partition 0 in the \gaia\ CAMD is shown in Fig.~\ref{fig:partitions}.}
    \label{fig:composita0}
\end{figure*}

\begin{figure*}
	\includegraphics[width=\textwidth/2, trim= 0mm 0mm 0mm 0mm]{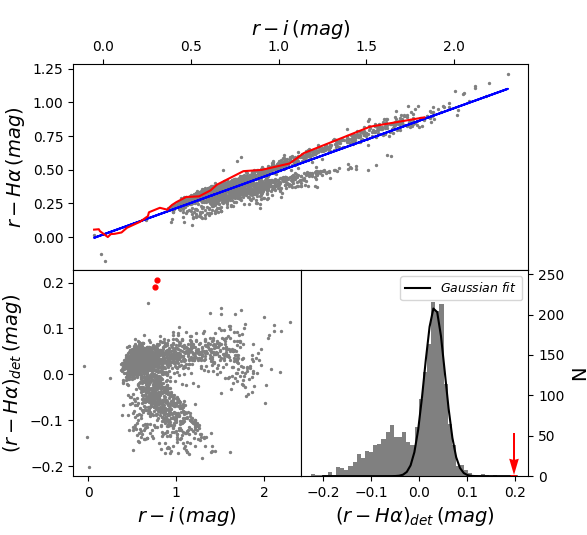}
    \caption{\normalsize Graphical depiction of the detrending and outliers selection processes performed on positional-partition 154. The red line in the top panel represents the synthetic ZAMS track, for zero reddening \protect\citep{drew}.}
    \label{fig:composita_pos154}
\end{figure*}

We point out that Chauvenet's criterion assumes an underlying Gaussian population, while a non-negligible amount of our partitions seems to deviate from this (mainly due to population mixing). However, the application of Chauvenet's criterion on non-Gaussian partitions provides a more robust \ha-excess outlier selection, since the standard deviation of these partitions is overestimated.
As can be noticed from the bottom-right panel of Fig.~\ref{fig:composita7331}, the detrended $r-$\ha\ distribution relative to the CAMD-partition 7331 constitutes a good example of Gaussian underlying population. 
On the other hand, Fig.~\ref{fig:composita0} presents an example of partition (CAMD-partition 0) in which the underlying distribution deviates from a standard Gaussian distribution.
As a reference, Fig.~\ref{fig:composita_pos154} presents the TCDs (before and after the detrending process) and the histogram of the detrended $r-$\ha\ values relative to positional-partition 154. Two trends are identifiable from the TCDs: the top one represents the locus in which unreddened MS stars lie, while the bottom trend corresponds to the reddened Red Giants (RG) track. These two trends reflect in the bimodality recognisable in the histogram in the bottom-right panel. This effect does not alter the number of outliers selected from partitions that present it, since the second Gaussian population is always redder than the main one.

Our algorithm selects both positive and negative outliers; however, since our goal is to identify the \ha-$\mathit{excess}$ candidates, the term ``outliers" will henceforth refer to only the positive ones.

\section{RESULTS}
\label{sec: results}

Our selection identifies 28,496 $r-$\ha\ 3$\sigma$ outliers (0.4\% of the total dataset) above the previously identified stellar loci. More specifically, 25,030 outliers are selected from the CAMD-partitions and 8,550 from the positional-partitions.

\begin{figure}
	\includegraphics[width=5\columnwidth/6, trim= 0mm 0mm 0mm 0mm]{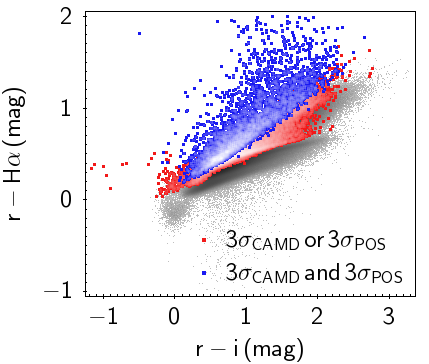}
	\includegraphics[width=5\columnwidth/6, trim= 0mm 0mm 0mm 0mm]{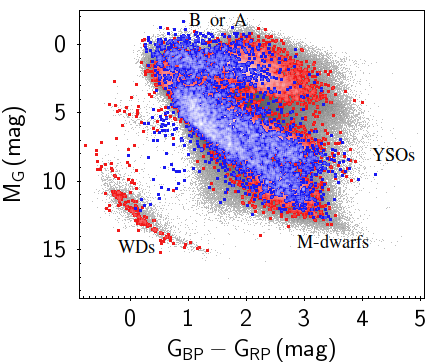}
    \caption{\normalsize The top panel shows the layout of the $r-$\ha\ $3\sigma$ outliers in the IPHAS TCD, while their position in the \gaia\ CAMD is presented in the bottom panel. The red dots represent all the $3\sigma$ outliers identified by either our CAMD-based or positional-based selection, while the blue dots represent the subset of 5,084 outliers selected from both the partition types. The intensity of both these colours scales inversely with the density of objects.}
    \label{fig:emittersCMD_CCD}
\end{figure}

In Fig.~\ref{fig:emittersCMD_CCD}, the locations in the IPHAS TCD and in the \gaia\ CAMD of these outliers are presented. It appears particularly noticable in the top panel that many of these candidates would have not stood out as outliers, if the chosen statistical analysis had been applied directly in the two-colours domain. From the bottom panel, mainly four regions of the CAMD with a particularly high density of outliers can be highlighted: the white dwarfs (WDs) track, the M-dwarfs area, the YSOs region, and the region where reddened early MS stars of spectral type B or A sit.

\begin{figure}
	\includegraphics[width=6\columnwidth/7, trim= 0mm 0mm 4mm 0mm]{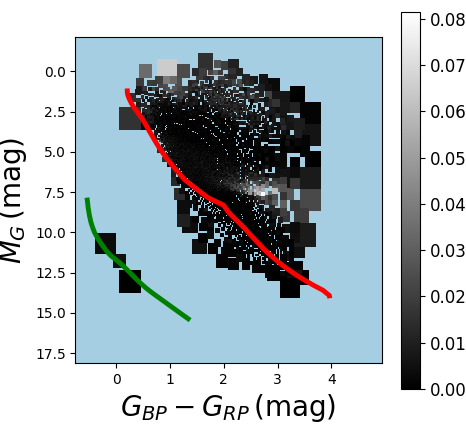}
	\caption{\normalsize Fraction of outliers per CAMD-partition.}
    \label{fig:emitters_fraction}
\end{figure}

Fig.~\ref{fig:emitters_fraction} displays the map of the fraction of outliers per CAMD-partition. Mainly two regions with a relatively high fraction of outliers stand out. These regions are: the area commonly associated to YSOs, i.e. to the right of the ZAMS (the red line) and centred at around $M_G = 7.5$\,mag, and the region where reddened, bright, B or A spectral types stars lie (i.e. the top area in the CAMD, centred around $G_{BP}-G_{RP} = 1.2$\,mag). Further information about the statistical composition of the \ha-excess candidates that occupy these areas of the CAMD can be obtained through a cross-match with SIMBAD database \citep{simbad}. Out of 981 outliers that occupy the former overpopulated region, 608 are classified as YSOs (or candidates), or T-Tauri stars; 154 of them are classified as Emission-line objects, while 146 simply as ``Star"\footnote{We point out that the generic ``Star" label in SIMBAD refers to objects that have been identified, but for which there is not enough information for a more specific classification.}. Among the outliers included in the latter group of interest, 131 find a classification in SIMBAD: 55 of them are identified Be stars (or candidates), 34 are Emission-line stars, 24 are classified as ``Star" and 10 are Red Giant Branch stars.

\begin{figure}
	\includegraphics[width=9\columnwidth/10, trim= 0mm 0mm 0mm 0mm]{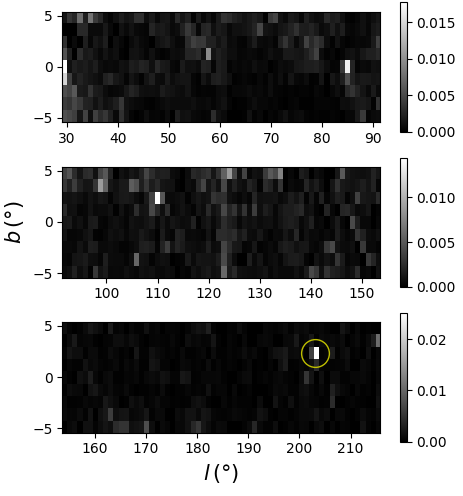}
	\caption{\normalsize Fraction of outliers per positional-partition. Positional-partition 1289 is highlighted by the yellow circle.}
    \label{fig:pos_emitters_fraction}
\end{figure}

\begin{figure}
	\includegraphics[width=5\columnwidth/6, trim= 0mm 0mm 0mm 0mm]{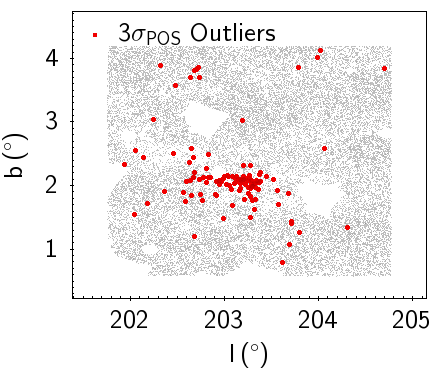}
	\caption{\normalsize Graphical depiction of the nine positional-partitions centred around positional-partition 1289 in the Galactic coordinates space. This latter partition is the one with the highest fraction of \ha-excess candidates (the red dots).}
    \label{fig:pos1289}
\end{figure}

In Fig.~\ref{fig:pos_emitters_fraction}, the map of the fraction of outliers per positional-partition is shown. To rule out systematic effects, an analysis of the relationship between the size of each partition and the fraction of outliers within it was performed; no such correlation was found. The distribution of this ratio in the Galactic coordinate space is consistent with being homogeneous, with no significant trend in either direction. Nonetheless, some areas in the $b$ vs. $l$ diagram with a relatively high density of \ha-excess candidates can be highlighted. These might correspond, say, to regions with a high rate of star formation, such as molecular clouds, or to open clusters. Two examples are the known open clusters IC1396 (centred at $l=99.30$\,\textdegree, $b=03.74$\,\textdegree; \citealp{IC1396}) and NGC2264 (centred at $l=202.94$\,\textdegree, $b=02.30$\,\textdegree; \citealp{NGC2264bis}, \citealp{NGC2264_coords}, \citealp{NGC2264}), which are easily identifiable in Fig.~\ref{fig:pos_emitters_fraction}. The latter star-forming region has been the subject of previous studies, such as the one presented by \cite{NGC2264}. They applied the method of the \textit{Bayesian inference} to identify 115 accreting objects in NGC2264. Positional-partition 1289 (highlighted by the yellow circle in Fig.~\ref{fig:pos_emitters_fraction}), corresponds to the sky area in which NGC2264 is located, and is in fact the positional-partition with the highest fraction of \ha-excess candidates: out of 2,826 objects, our algorithm selects 71 outliers (2.5\%). Nine positional-partitions centred around this partition are shown in Fig.~\ref{fig:pos1289}. The apparently empty areas in the sky are due to the quality cuts applied when compiling IPHAS DR2; because of these cuts, IPHAS DR2 provides photometric measurements for sources covering $92\%$ of its footprint \citep{barentsen}.

\smallskip

Ideally, the concept of ``outliers of a distribution" would be non-arbitrary. However, realistically speaking the definition of ``outlier" is strongly dependent on the chosen threshold. This can be mitigated by the choice of different confidence levels during the selection process. One possibility consists in considering as outliers all the objects that are selected using Chauvenet's criterion: this would be ideal if all partitions were to display Gaussian distributions. On the other hand, one can choose to select as outliers all the objects that satisfy $\sigma\geq 3$ (Eq.~\ref{eq:selection}); this constitutes a more relaxed threshold, when compared to Chauvenet's one. We point out that by setting this threshold, a certain amount of false-positives in our selection is to be expected. However, this amount is not easily quantifiable, since the distributions are not always Gaussian, and also they are not equally populated. The suggested threshold is the 5$\sigma$ one ($\sigma\geq 5$), as a compromise to reduce false-positives fraction, while retaining a robust candidate selection. In fact, all the candidates selected using a 5$\sigma$ threshold would have been included using Chauvenet's criterion as well. By applying the 5$\sigma$ cut, 6,774 outliers (0.09\% of the complete dataset, 23.8\% if compared to the 3$\sigma$ sample) are identified: 6,455 from the CAMD-partitions and 2,209 from the positional-partitions.

For all the sources in the master-catalogue, the two $flagCAMD$ and $flagPOS$ specifications are evaluated. These entries can assume a value of 0 (if the significance is lower than 3), 1 (if the significance is greater than or equal to 3, but smaller than 5), or 2 (if the significance is equal to or greater than 5). A very similar classification of the significance levels was previously adopted by \cite{witham} and by \cite{monguio}.

\smallskip

Our results are presented in a \textit{meta-catalogue}, the first 10 rows of which are shown in Table~\ref{tab:light_catalogue}. 
The full set of metrics computed during the catalogue generation is also published. In this full version, the necessary pieces of information to trace back each source to the corresponding CAMD-based and positional-based partitions are provided, as well as the detrending model information for each TCD. Our hope is that this additional information will aid future users of the catalogue to further tune the selection of \ha-excess sources to suit a specific task.


\section{DISCUSSION}
\label{sec: discussion}

\begin{figure*}
	\includegraphics[width=5\columnwidth/6, trim= 0mm 0mm -3mm 0mm]{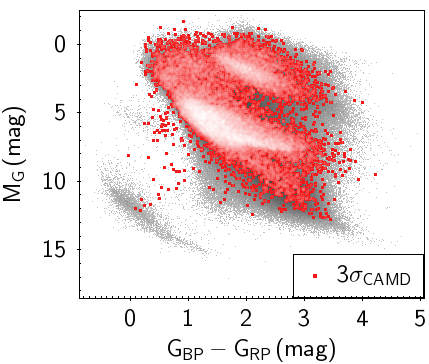}
	\includegraphics[width=5\columnwidth/6, trim= -3mm 0mm 0mm 0mm]{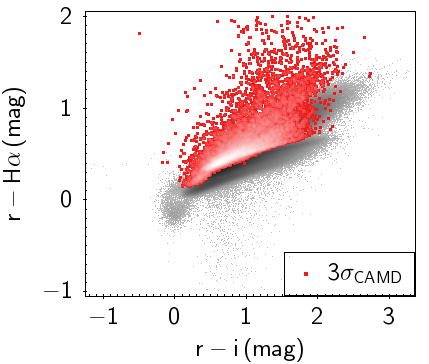}
	\includegraphics[width=5\columnwidth/6, trim= 0mm 0mm -3mm 0mm]{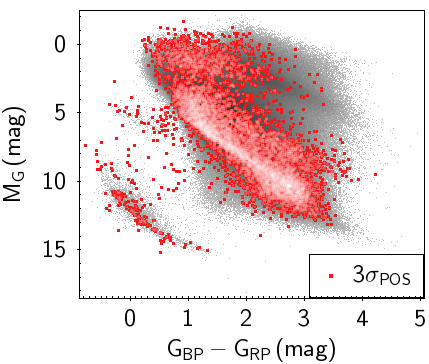}
	\includegraphics[width=5\columnwidth/6, trim= -3mm 0mm 0mm 0mm]{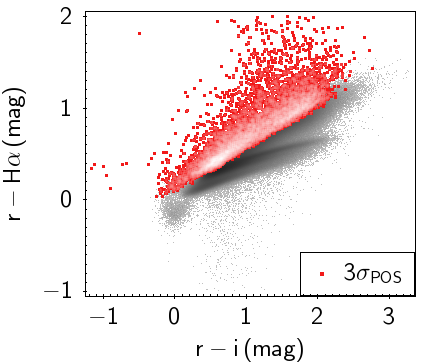}
    \caption{\normalsize The top row presents the $r-$\ha\ $3\sigma$ outliers (the red dots) identified from CAMD-partitions, while the ones obtained from positional-partitions are shown in the bottom row.}
    \label{fig:significances}
\end{figure*}

In Fig.~\ref{fig:significances} the positions of our $3\sigma$ outliers in the CAMD (left column) and in the TCD (right column) are shown.
The most evident differences between the selections applied on the CAMD-partitions and on the positional-partitions are:
\begin{enumerate}
    
    \item The CAMD-based selection is less efficient, compared to the positional-based one, in identifying outliers along the WD track. This effect stands out clearly from the comparisons of both the TCDs and the CAMDs. It is due to the constraints set when partitioning the \gaia\ CAMD: the partitions in the least populated regions of the CAMD have to be large enough in size to contain at least 500 sources each. For this reason, our algorithm creates only three large partitions in the WD track and in the surrounding area of the CAMD. This results in a limited amount of detected outliers.

    \item The CAMD-based selection identifies more \ha-excess candidates in the region of the CAMD where the reddened B and A types emission line stars are, if compared to the positional-based algorithm.
    
    \item Most M-dwarf \ha-excess candidates are identified mainly through the positional-based selection. We believe that this is related to the fact that M-dwarf stars have an intrinsically more intense $r-$\ha\ IPHAS colour, compared to other MS stars. In contrast with the CAMD-partitions, in the positional-partitions these objects are blended with other populations, and hence they stand out as outliers.
    
    \item The CAMD-based selection is more efficient at identifying YSOs of various types. This can be observed in the top CAMD in Fig.~\ref{fig:significances} as the cluster of \ha-excess candidates found to the right of the MS track, and centred at around $M_G\sim7.5$\,mag. These systems would be difficult to identify with a positional-based partition, unless they display strong \ha\ emission.

\end{enumerate}

The position of the sources in the CAMD constitutes an indication of the stellar population they most likely belong to. A cross-match between our outliers sample and the SIMBAD database \citep{simbad} provides a further statistical representation of the populations our \ha-excess candidates belong to. A cross-matching radius of 1\,arcsecond yields 1,825 matches. Of them, 822 are classified as YSOs (or candidate YSOs) or T-Tauri stars, 376 sources are classified with the generic epithet of ``Star", 233 are Emission-line stars, 113 are classified as Be stars (or candidates), 44 as Orion Variable stars, 13 as WDs (plus 43 WD candidates), and 8 are known CVs (or candidates). The WDs included in our list of outliers and in SIMBAD are further classified, according to their spectral type: six of them are DB white dwarfs, four are DA type, two DC type, one DAB type and one DBA type. The non-DA type WDs are identified as \ha-excess candidates by our algorithm because they do not present the strong absorption lines, typical of DA type WDs.

\smallskip

\begin{figure*}
	\includegraphics[width=5\columnwidth/6, trim= 0mm -5mm -3mm 0mm]{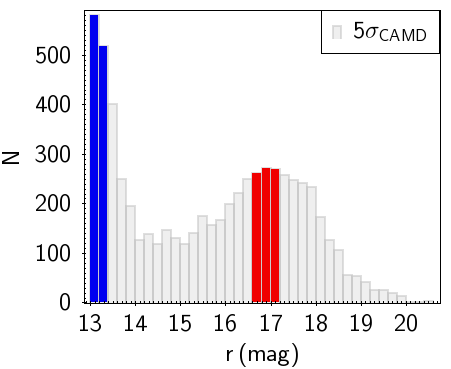}
	\includegraphics[width=5\columnwidth/6, trim= -3mm -5mm 0mm 0mm]{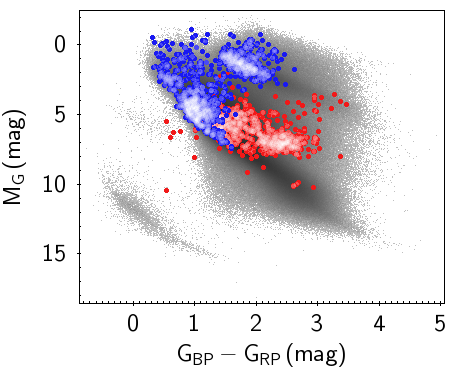}
	\includegraphics[width=5\columnwidth/6, trim= 0mm -5mm -3mm 0mm]{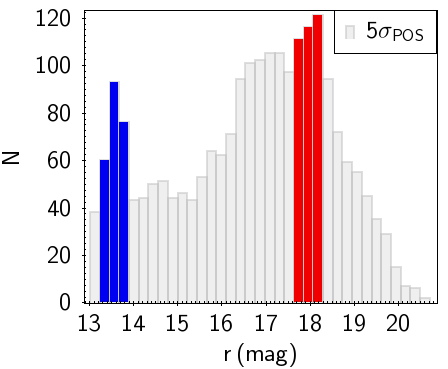}
	\includegraphics[width=5\columnwidth/6, trim= -3mm -5mm 0mm 0mm]{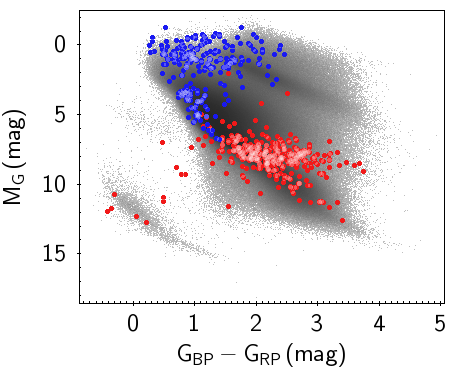}
    \caption{\normalsize $r$ magnitude
    distributions (left column) of the $5\sigma_{CAMD}$ outliers (top row) and of the $5\sigma_{POS}$ outliers (bottom row), with a bin width of 0.2\,mag. Both the distributions are bimodal; the blue areas in the histograms point to the bins around the brightest of the two modes, respectively, while the red areas in the histograms display the three most populated bins around the secondary modes. In the right column, the density in the \gaia\ CAMD of the objects that occupy the areas around the modes in the corresponding histogram are shown.}
    \label{fig:color_histograms}
\end{figure*}

The left column in Fig.~\ref{fig:color_histograms} shows the $r$ magnitude distributions of our $5\sigma$ outliers. The bin size for these histograms is 0.2\,mag. As can be noticed, both the distributions are bimodal; the peak around the 13th magnitude for the CAMD-outliers, as well as the one around $r=13.5$\,mag for the positional-outliers, is to be partially imputed to an observational bias. The secondary mode for the $5\sigma_{CAMD}$ outliers is 16.90\,mag, and it is very close to the mode of the \textit{r} intensity of the whole data set (which is $r\sim 16.60$\,mag). On the other hand, the most frequent \textit{r} intensity for the $5\sigma_{POS}$ outliers is 18.15\,mag (i.e. more than 1.5 magnitude fainter than the mode of the whole data set), confirming the fact that, generally speaking, our positional-selection is more efficient in identifying fainter outliers than the CAMD-selection. The blue areas in the histograms represent the most populated bins around $r =13$\,mag (for the CAMD-outliers) and $r =13.5$\,mag (for the positional-outliers), while the red areas indicate the three most populated bins around the respective secondary modes. In the right column of the same figure, the CAMD densities of the sources belonging to these coloured regions are shown. According to their positions in the CAMD, the brightest outliers in both the distributions are active B or A types stars. The CAMD-outliers belonging to the red bins mainly cluster in the region of the CAMD associated to YSOs. Also the positional-outliers with an r magnitude close to the secondary mode mainly occupy the region of the CAMD associated to YSOs; however, some of them lie on the WD track, some in between the WD track and the MS (making them good CV candidates), and some can be found on the M-dwarfs region.

\bigskip

In the following subsections, our results are compared with previous similar studies. More specifically, they are cross-matched with the catalogues developed by \citealp{witham} and by \citealp{monguio} (IGAPS). Moreover, a further validation of our selection, based on the visual inspection of LAMOST DR5 spectra \citep{2019lamost} is presented.
Since accreting compact objects often show X-ray emission, a fraction of our \ha-excess candidates are expected to be found in X-ray surveys as well. Therefore, in the last subsection, the quantitative results of the cross-matches with three X-ray surveys are briefly discussed. These surveys are: the \textit{ROSAT} All-Sky Survey Faint Source Catalogue (``faint-\textit{ROSAT}" from now on; \citealp{faint_rosat}), the \textit{ROSAT} All-Sky Survey Bright Source Catalogue (``bright-\textit{ROSAT}" from now on; \citealp{bright_rosat}) and the \textit{Chandra} Source Catalogue (``CSC" from now on; \citealp{chandra_bis}).

\subsection{Comparison with Witham's catalogue}
\label{sec: comparison_witham}

Comparing the emitters' list in \citealp{witham} (which counts 4,853 objects) with the full master-catalogue developed by \citealp{Scaringi} (after the application of the quality cuts described in Sec.~\ref{sec:cuts}), 1,213 common sources (25.0\% of \citealp{witham} outliers sample) are found. The cross-match is performed by using a radius of 1\,arcsecond; however, the number of matches does not change significantly if this parameter is increased up to a generous 5\,arcseconds. Although the remaining 75\% of Witham's outliers can be found in \gaia\ DR2 archive, their astrometric/photometric measurements did not satisfy the quality constraints applied by \cite{Scaringi} when producing the \gIPHAS.

\begin{figure}
	\includegraphics[width=5\columnwidth/6, trim= 0mm -4mm 0mm 0mm]{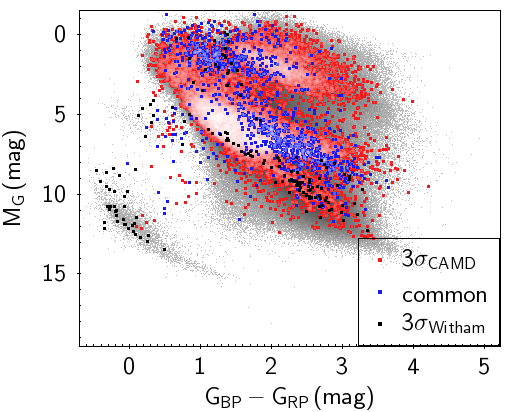}
	\includegraphics[width=5\columnwidth/6, trim= 0mm 0mm 0mm -4mm]{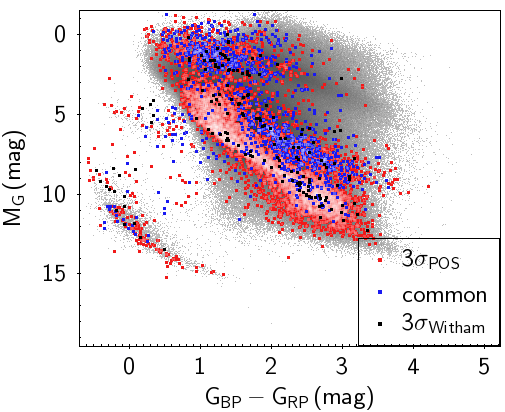}
    \caption{\normalsize Top panel: position in the CAMD of the matches between the outliers in \protect\cite{witham} and our $3\sigma$ CAMD-outliers.
    Bottom panel: position in the CAMD of the matches between the outliers in \protect\cite{witham} and our positional-outliers (bottom-panel). The red dots represent the totality of our CAMD/positional outliers; the blue dots are the common outliers between Witham's list and our CAMD/positional selection; the black dots are Witham's outliers not selected by our CAMD/positional algorithm.}
    \label{fig:cross_witham}
\end{figure}

Out of the 1,213 common targets, 1,115 (91.9\%) are identified as outliers by our algorithm as well, with a significance (either $\sigma_{CAMD}$ and/or $\sigma_{POS}$) equal to or higher than 3. By comparing the common sources with our CAMD-based outliers, 1,053 of common outliers (94.4\%) are recovered, whilst the positional-based selection finds 1054 (94.5\%) of them (i.e. 992 common outliers are identified by both our selection criteria).
A subset of 933 out of 1,115 common objects is characterised by $\sigma_{CAMD}\geq 5$ and/or $\sigma_{POS} \geq 5$; 893 of them have a $\sigma_{CAMD}\geq 5$, while 671 of them have $\sigma_{POS} \geq 5$ (hence 631 Witham's emitters are found by both our selection criteria, with a significance equal to or greater than 5). In the two panels of Fig.~\ref{fig:cross_witham}, the positions in the \gaia\ CAMD of the objects resulting from the cross-match with Witham's catalogue are presented. More specifically, the top-panel shows the matches between Witham's list and our $3\sigma$ CAMD-outliers, while the bottom-panel shows an analogous diagram for our positional-outliers. As previously stated in this work, the CAMD-based selection is more efficient in recovering bright objects, such as the ones that lie on the active, reddened, B or A types stars track. On the other hand, the cross-match with our positional-based selection yields more matches in the M-dwarf region of the CAMD, and on the WD track.

\smallskip

\begin{figure}
	\includegraphics[width=5\columnwidth/6, trim= 0mm 0mm 0mm 0mm]{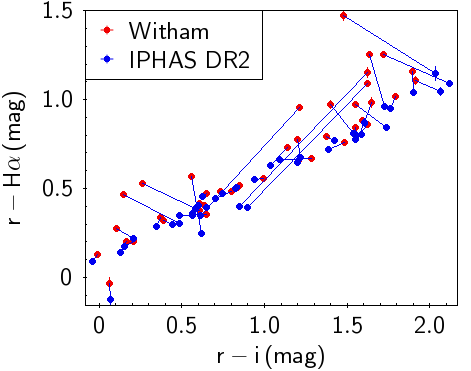}
	\caption{\normalsize Positions in the TCD of 42 sources that were identified as outliers in \protect\cite{witham}, but not my our algorithm. These are the objects the $r-$\ha\ intensity of which was higher in Witham's catalogue (the red dots) with respect to the analogous IPHAS DR2 intensity (the blue dots). For most of these targets, the error bars included in the plot are too small to be visible.}
    \label{fig:dr1vsdr2}
\end{figure}

By checking the differences in the photometric measurements in \citealp{witham} (minus the uncertainties) and in IPHAS DR2 (plus the uncertainties), some variable objects can be spotted. However, we point out that this apparent variability might be due to the different calibrations\footnote{As previously mentioned, the study of \cite{witham} was performed on IPHAS pre-publication measurements, while our selection algorithm is applied to IPHAS DR2 calibrated data.}. Of the 98 3$\sigma$ objects missing in our list of outliers, 42 showed a stronger $r-$\ha\ emission at the epoch of Witham's study, if compared to IPHAS DR2. This decrease in the $r-$\ha\ intensity might be the reason why those particular sources were identified as \ha\ emitting candidates in \cite{witham}, but not by the methods implemented in our study. Fig.~\ref{fig:dr1vsdr2} shows this apparent $r-$\ha\ intensity drop.

\subsection{Comparison with IGAPS}
\label{sec: comparison_IGAPS}

The identification of emission-line objects performed by \cite{monguio} followed a selection strategy that is similar to the one implemented for \cite{witham}. The main differences between these two works are related to the data calibration and to the morphology classes being tested (\citealp{monguio} only excluded ``morphology class 0" sources, i.e. the ``noise-like sources", from being tested for \ha\ excess; see also \citealp{farnhill}). Of the 53,234,833 objects in the IGAPS catalogue that were tested for \ha\ excess, \cite{monguio} produced a list of 20,860 excess-line candidates (0.04\% of the tested targets). These outliers were selected with a significance higher than 3; a sub-sample of these excess candidates is composed by 8,292 objects (0.02\% of the tested sample) with significance higher than 5.

\begin{figure*}
	\includegraphics[width=2\textwidth/3, trim = 0mm 0mm 0mm 0mm]{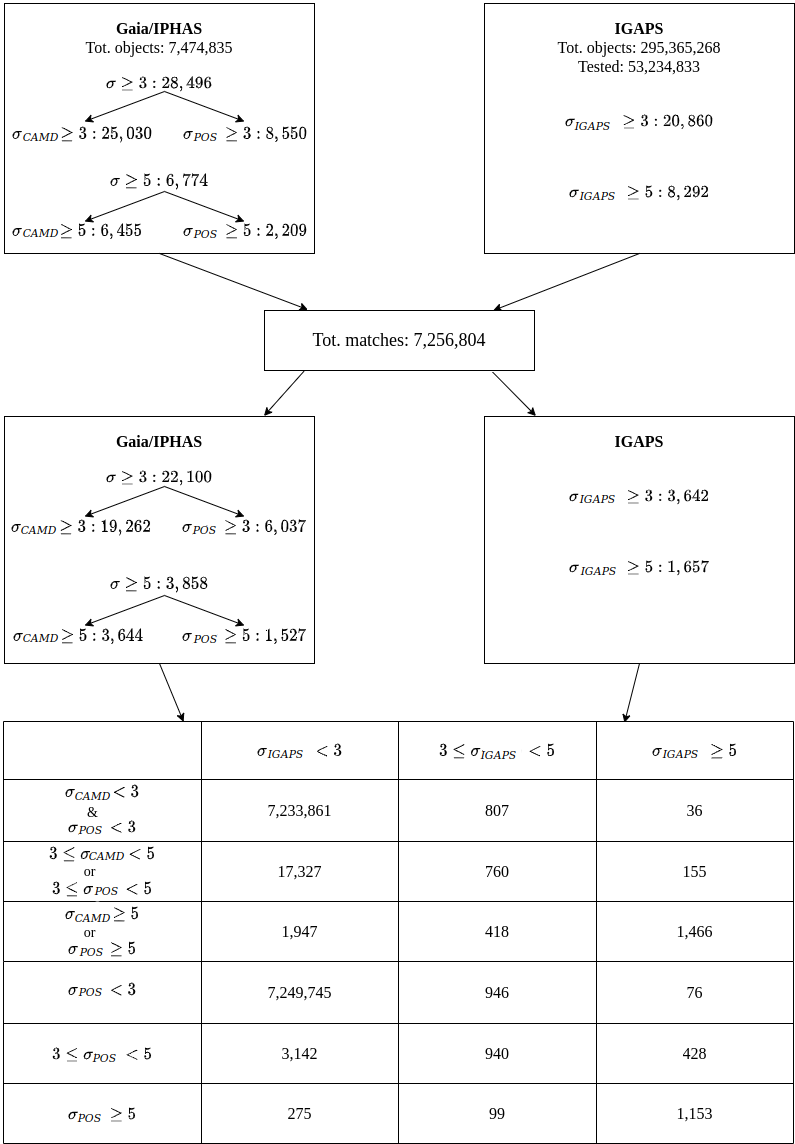}
    \caption{\normalsize The flow-chart describes the cross-matching process between the \gIPHAS\ and the IGAPS catalogue, as well as its detailed results.}
    \label{fig:cross_igaps}
\end{figure*}

A cross-match between the \gIPHAS\ (after the application of the quality cuts described in Sec.~\ref{sec:cuts}) and the $\sim53$\,million IGAPS tested sources yields 7,256,804 matches. The cross-matching radius is 1\,arcsecond. This subset includes 3,642 IGAPS outliers, 1,657 of which with an associated significance higher than 5. It also includes a subset of 22,100 of our 3$\sigma$ outliers: 19,262 of them derive from the CAMD-based selection, and 6,037 from the positional-partitions.
We point out that our CAMD-based partitioning algorithm is performed on a different parameter space with respect to the one applied in \cite{monguio}. Nonetheless, for a more complete discussion, all the results of the possible cross-matches between the two lists are provided. The cross-matching process between these two catalogues and its results are presented in the flow-chart in Fig.~\ref{fig:cross_igaps}.

\begin{figure*}
	\includegraphics[width=5\columnwidth/6, trim= 0mm 0mm -4mm 0mm]{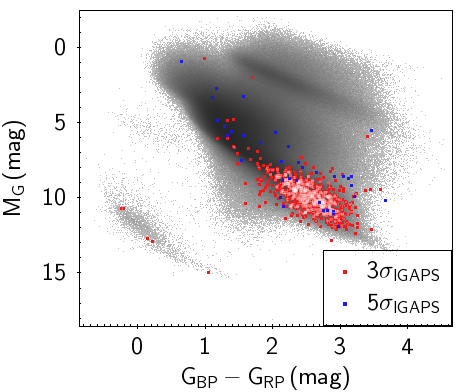}
	\includegraphics[width=5\columnwidth/6, trim= -4mm 0mm 0mm 0mm]{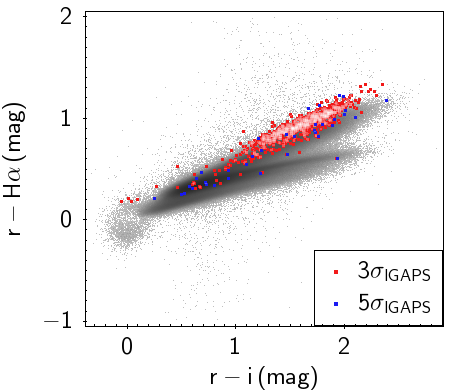}
	\includegraphics[width=5\columnwidth/6, trim= 0mm 0mm -4mm 0mm]{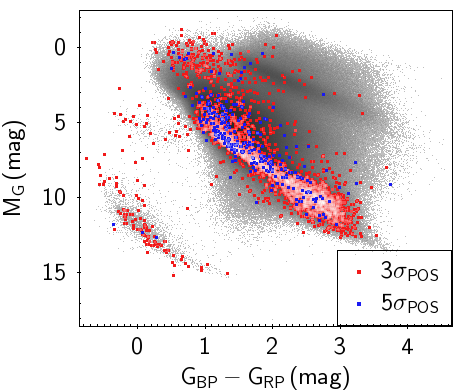}
	\includegraphics[width=5\columnwidth/6, trim= -4mm 0mm 0mm 0mm]{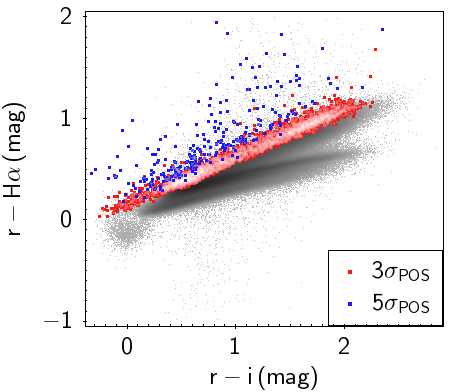}
    \caption{\normalsize Top row: location in the \gaia\ CAMD (left panel) and in the IPHAS TCD (right panel) of the 843 IGAPS $r-$\ha\ outliers that are not selected by our algorithm. The red dots represent objects that \protect\cite{monguio} identified with a significance included between 3 and 5, while the objects with a higher significance are depicted with blue dots.
    Bottom row: position in the CAMD (left panel) and in the TCD (right panel) of our 3,417 positional-outliers that are not listed as \ha-excess candidates in IGAPS. The red dots represent the positional-outliers with a significance included between 3 and 5, while the blue dots represent outliers with a higher positional-significance.}
    \label{fig:missing_igaps_map}
\end{figure*}

The positions in the \gaia\ CAMD and in the IPHAS TCD of the 843 IGAPS outlier not identified by our algorithm are shown in the top row of Fig.~\ref{fig:missing_igaps_map} (the red and blue dots). IPHAS DR2 photometric measurements are used to produce the TCD. As can be noticed from the CAMD, the vast majority of these objects can be associated to the M-dwarf region of the CAMD. As previously mentioned, these objects are characterised by significantly different IPHAS colours, with respect to the other MS stars. This appears to fail our selection through the use of CAMD-based partitions.
In the bottom row of the same figure, our 3,417 positional-outliers that were not identified by \cite{monguio} are placed in the CAMD and in the TCD. These objects mainly lie in the CAMD on the MS track, on the reddened B and A types stars track, or on the WD track. However, some red dots are placed in between these two tracks, making them good CV candidates.

\begin{figure}
	\includegraphics[width=6\columnwidth/7, trim= 0mm 0mm 0mm 0mm]{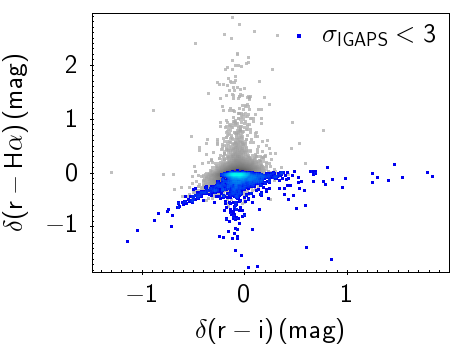}
    \caption{\normalsize Position in the $\delta$\,($r-$\ha) vs. $\delta$\,($r-i$) diagram of all the matches between IGAPS and the \gIPHAS\ (the grey dots), and of our positional-outliers that \protect\cite{monguio} did not select as \ha-excess candidates (the overplotted blue dots).}
    \label{fig:calibration}
\end{figure}

The mismatches between the results obtained in \cite{monguio} and by us are to be ascribed mainly to two factors: the different definitions used to calculate $\sigma$ and the different calibrations applied to the photometric measurements in the input databases. In fact, when producing the \gIPHAS, \cite{Scaringi} based their work on IPHAS DR2 calibrated data, while IGAPS calibration (as mentioned in section~\ref{sec: introduction}) relies on the more recent ``Pan-STARRS reference ladder" \citep{magnier}.
This latter effect is visible in Fig.~\ref{fig:calibration}, where the $\delta$\,($r-$\ha) vs. $\delta$\,($r-i$) diagram is presented. The two axes represent the difference between IGAPS and IPHAS DR2 values for $r-$\ha\ and $r-i$ parameters, respectively. The grey dots correspond to all the matches between IGAPS and the \gIPHAS, while the blue dots are our positional-outliers that \cite{monguio} did not identify as \ha-excess candidates. The $\delta$\,($r-i$) mode for the grey dots is -0.05\,mag, while the most common value for the blue dots is -0.06\,mag. The mode of the $\delta$\,($r-$\ha) distribution for both the grey and blue points is -0.03\,mag. If IGAPS and IPHAS DR2 data had the same calibration, the points in this diagram would cluster around the (0,0) coordinates. However, almost all our positional-outliers not listed in IGAPS lie below the $\delta$\,($r-$\ha)=0 line; this supports our hypothesis that the different calibration is one of the main factors that cause the discrepancy between our positional-selection and IGAPS selection.

\subsection{LAMOST spectra\protect\footnote{In the current section, we refer to the sources using their LAMOST DR5 designations, as well.}}

A more direct validation of our selection comes from visually inspecting the spectra of the photometrically identified \ha-excess candidates. In order to achieve this validation, a cross-match between our list of outliers with LAMOST DR5 is performed. However, LAMOST DR5 spectra and IPHAS DR2 measurements were acquired at different epochs (IPHAS DR2 observations were implemented between 2003 and 2012, while LAMOST DR5 spectra were collected between 2016 and 2017). Therefore some transient \ha-excess sources selected by our algorithm may not display clear \ha\ emission, and vice-versa.

\subsubsection{Purity and completeness}

\begin{figure*}
	\includegraphics[width=\columnwidth, trim= 0mm -7mm 0mm 0mm]{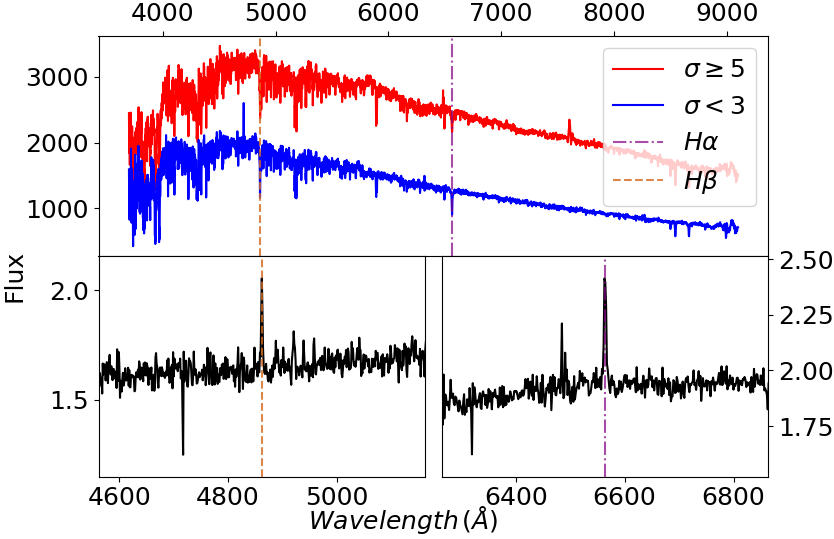}
	\includegraphics[width=\columnwidth, trim= -4mm -7mm 0mm 0mm]{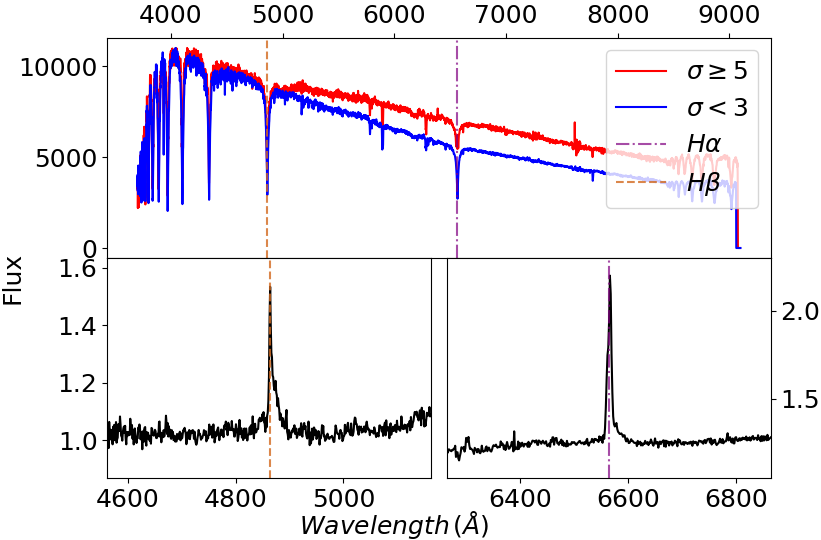}
	\caption{\normalsize Example of two $5\sigma$ \textit{excess} sources that show \ha\ absorption (the red lines in the top panels). The blue lines represent the fluxes of two objects in the same partitions, with a significance lower than 3. The bottom panels display the ratios between the the fluxes in the top panels, centered around the H$\beta$ wavelength (bottom-left panels) and around the \ha\ wavelength (bottom-right panels).}
    \label{fig:absorb_less}
\end{figure*}

A cross-match with the LAMOST DR5 archive and our \ha-excess candidates list (with a 0.5\,arcsec cross-matching radius) yields 1,873 spectra. These spectra are used to calculate the \textit{purity} of our selection, with the assumption that they constitute a good representation of our $3\sigma$ outliers. Of these 1,873 objects, 916 (48.9\%) are confirmed as reliable \ha-excess candidates, while 939 (50.1\%) seem to show \ha\ absorption. The remaining 18 spectra do not allow a univocal assessment, due to their low quality. 
We point out that these relatively low spectral confirmation rates constitute a lower limit for the purity of our selection, since our algorithm does not aim to identify \ha\ \textit{emitters}, but rather \ha-\textit{excess} sources. Therefore, objects that exhibit excess \ha\ flux (but not necessarily displaying an \ha\ emission line) relative to the underlying partition are selected as outliers. This also explains the higher spectral confirmation rate for the positional-outliers, with respect to the CAMD-outliers. 
Fig.~\ref{fig:absorb_less} displays two examples of $5\sigma$ outliers the LAMOST spectra of which show absorption in the \ha\ band. These are compared to the spectra of two other objects in the same partitions with an associated significance lower than 3. The ratios between the red and blue fluxes in the top panels, zoomed in around the \ha\ wavelength, are presented in the bottom-right panels. In correspondence with the \ha\ wavelength, both these ratios are significantly above the mean, which explains the high significance associated to these sources. \ha\ excess is often accompanied by H$\beta$ excess, as can be seen in the bottom-left panels.

Purity does not change significantly, if a more conservative cut on the selection of the outliers is considered: out of 616 spectra relative to sources with either $\sigma_{CAMD}\geq 5$ and/or $\sigma_{POS}\geq5$, 306 (49.7\%) seem to be solid \ha-excess candidates. On the one hand, out of the 603 $5\sigma$ CAMD-outliers for which LAMOST spectra are available, 294 (48.8\%) show \ha-line emission. On the other hand, 128 spectra out of 157 (81.5\%) seem to validate our $5\sigma$ positional-based selection.
By applying a more rigid cut on IPHAS magnitudes, and hence reducing the effects due to saturation, the ratio of spectroscopically confirmed outliers improves significantly. In fact, retaining the sources with $r \geq 13.5$\,mag, $i \geq 12.5$\,mag and \ha $\geq 13$\,mag, 772 spectra out of our 1,141 (67.7\%) confirm our 3$\sigma$ outlier selection (either CAMD-based and/or positional-based). Constraining the spectral analysis to our 5$\sigma$ outliers, 231 LAMOST spectra out of 282 (81.9\%) confirm our selection. Thus, these additional quality-cuts are suggested to the users of our meta-catalogue.

\begin{figure}
	\includegraphics[width=6\columnwidth/7, trim= 0mm 0mm 0mm 0mm]{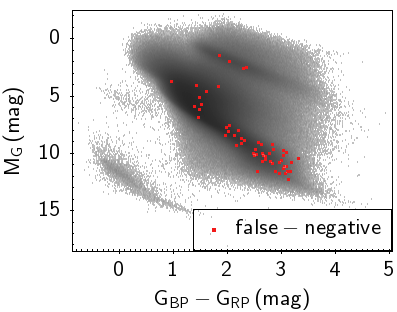}
	\includegraphics[width=6\columnwidth/7, trim= 0mm 0mm 0mm 0mm]{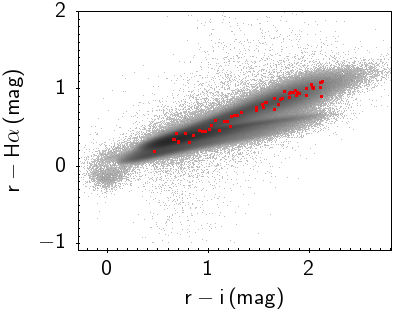}
    \caption{\normalsize Position in \gaia\ CAMD (top panel) and in IPHAS TCD (bottom panel) of 56 ``false-negative" objects.}
    \label{fig:false_negative}
\end{figure}

The \textit{completeness} parameter ($C$) relative to our selection (i.e. the ratio between the number of spectroscopically confirmed \ha\ emitters identified by our algorithm and the total amount of spectroscopically confirmed \ha\ emitters within our full master-catalogue) is obtained with two different methods:
\begin{itemize}
    \item the first method consists of the evaluation of:
    \begin{equation}
        C= \frac{N_{\sigma\geq3} \cdot P}{N_{\sigma\geq3} \cdot P + N_{\sigma<3} \cdot fn}
    \end{equation}
    Here, $N_{\sigma \geq/< 3}$ is the amount of objects with a significance higher/lower than 3, \textit{P} is the purity fraction relative to the full $3\sigma$ outliers sample ($48.9\%$), and \textit{fn} is the \textit{false-negative} fraction ($5.6\%$). This latter parameter derives from the visual inspection of 1,000 spectra belonging to randomly selected sources with $\sigma<3$, and corresponds to the fraction of these spectra that show \ha\ emission. This method yields a completeness of around $3\%$. The positions of the 56 false-negative objects in the CAMD and in the TCD are shown in Fig.~\ref{fig:false_negative} (top panel and bottom panel, respectively). Most of these sources lie in the M-dwarfs region of the CAMD;
    \item the second method consists of the visual inspection of 2,000 LAMOST spectra belonging to randomly selected objects in our catalogue, 15 if which are identified as \ha\ $3\sigma$ outliers by our selection. The completeness thus obtained is approximately 5\%.
\end{itemize}
Such low completeness values are partially due to the combination of a) the different epochs between LAMOST DR5 spectra and \gaia/IPHAS measurements (and hence the variability of some objects), b) too conservative thresholds during the \ha\ outliers selection processes, and c) too generous definition of ``\ha\ emitters" during the visual inspection of the spectra. Calibration-related problem are ruled out by the fact that none of our false-negative objects are included in IGAPS list of outliers. However, an exhaustive explanation for this low completeness fractions is still to be found. As a comparison, the same calculations applied on IGAPS catalogue yield a completeness percentage below 1\%.

\begin{figure*}
	\includegraphics[width=\columnwidth, trim= 0mm -7mm 0mm 0mm]{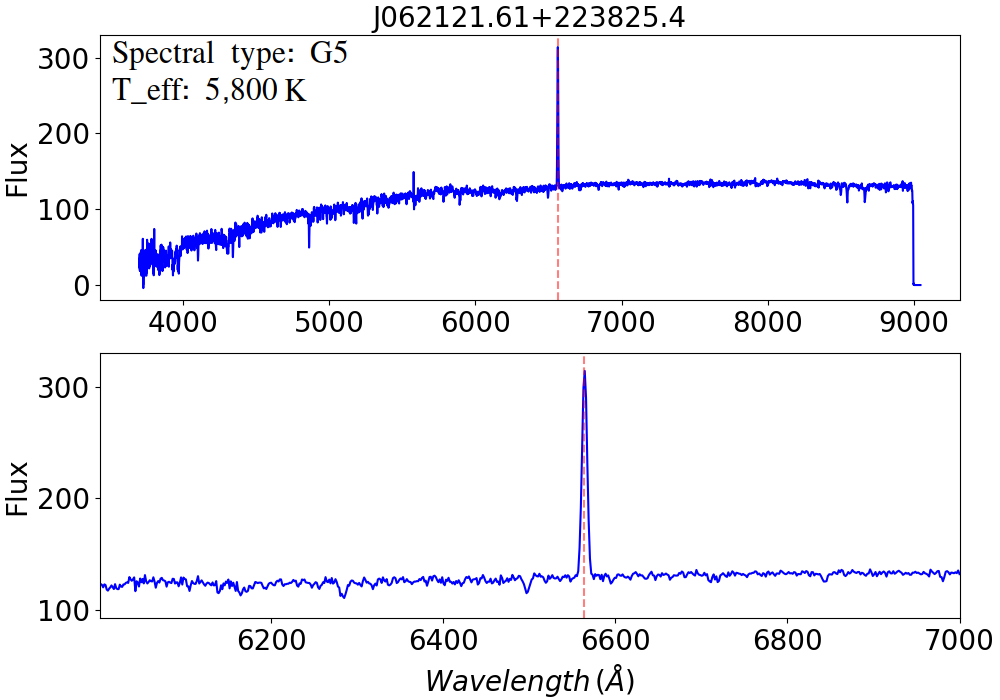}
	\includegraphics[width=\columnwidth, trim= 0mm -7mm 0mm 0mm]{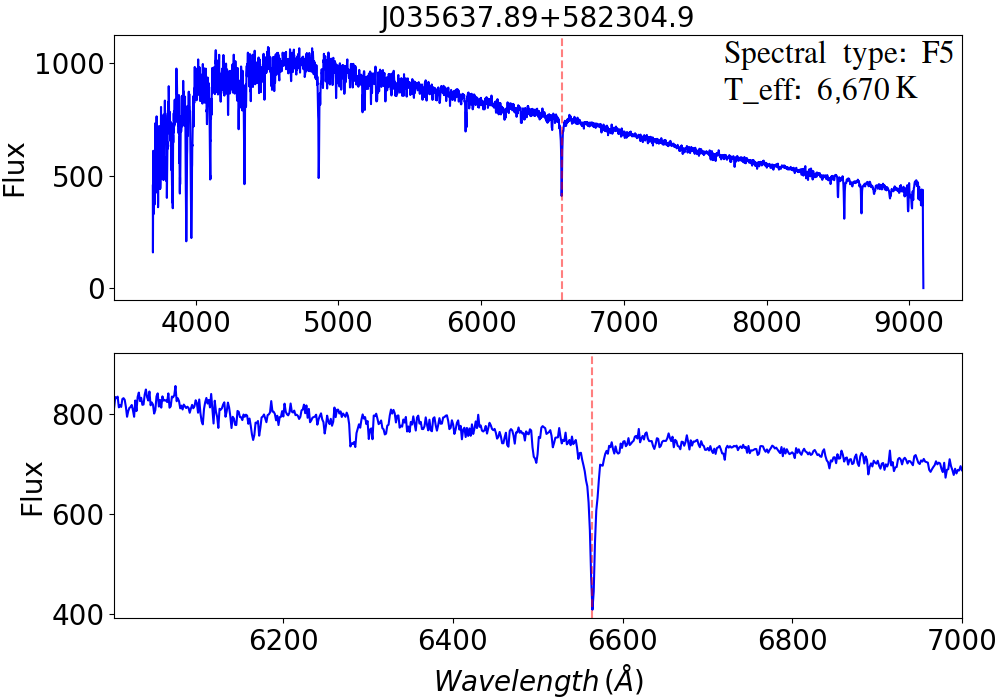}	\includegraphics[width=\columnwidth, trim= 0mm 3mm 0mm -7mm]{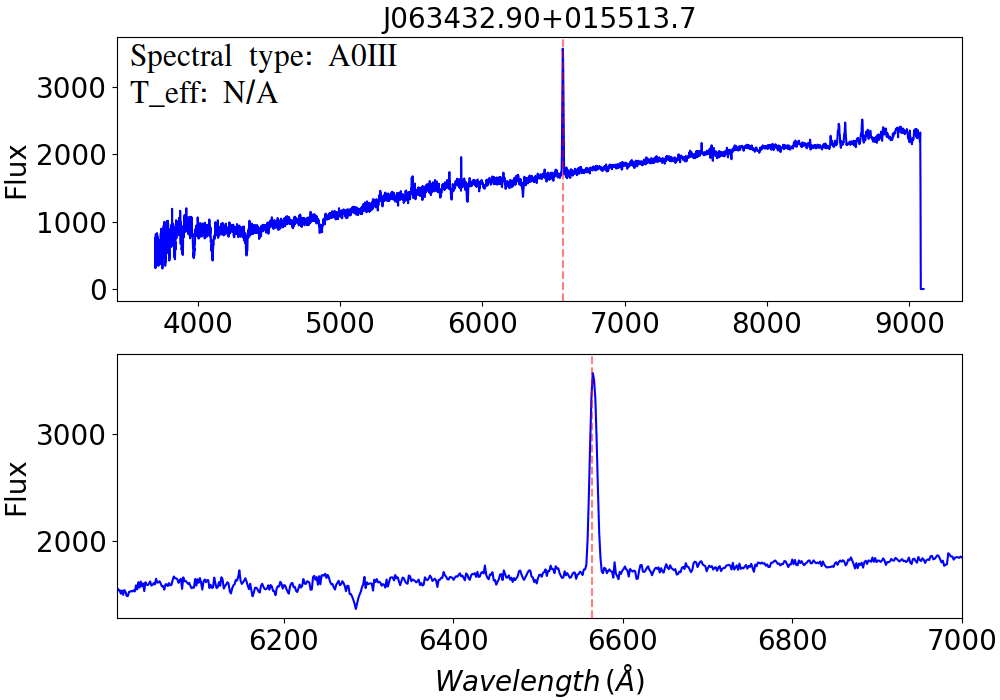}
	\includegraphics[width=\columnwidth, trim= 0mm 3mm 0mm -7mm]{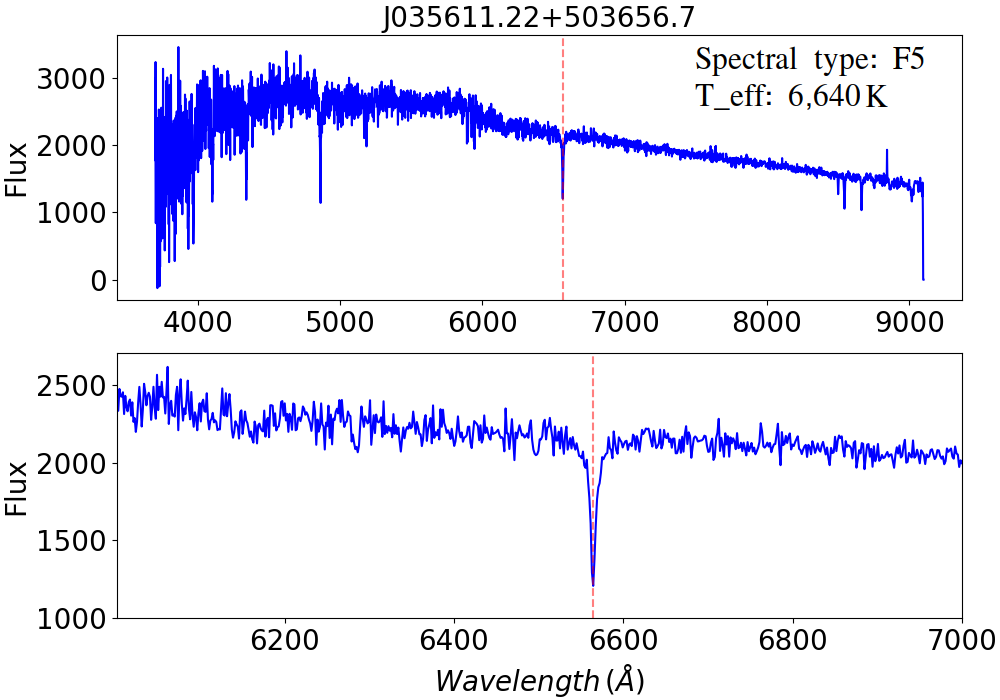}
	\caption{\normalsize LAMOST spectra of four objects that are selected as \ha-excess sources by our algorithm. For each of these spectra, a zoom to the region around the \ha\ line is shown. The top two spectra belong to the sources J062121.61+223825.4 (\gaia\ DR2 ID 3377220715714066304) and J035637.89+582304.9 (\gaia\ DR2 ID 470024698144186112) respectively, which have an associated significance (either CAMD-based and/or positional-based) included between 3 and 5. The bottom spectra refer to the objects J063432.90+015513.7 (\gaia\ DR2 ID 3120920947508407808) and J035611.22+503656.7 (\gaia\ DR2 ID 250435321081819392), which are characterised by a higher significance. The red dashed line indicates the \ha\ wavelength.}
    \label{fig:general_spectra}
\end{figure*}

\begin{figure*}
	\includegraphics[width=5\columnwidth/6, trim= 0mm 3mm 0mm -3mm]{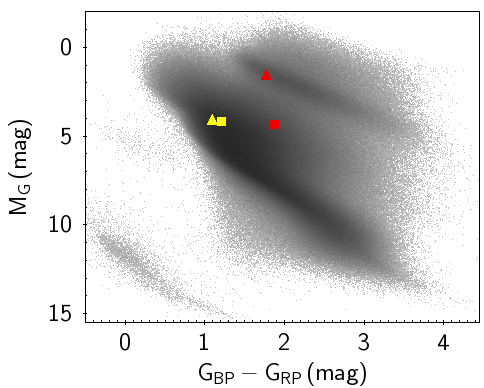}
	\includegraphics[width=5\columnwidth/6, trim= 0mm 3mm 0mm -3mm]{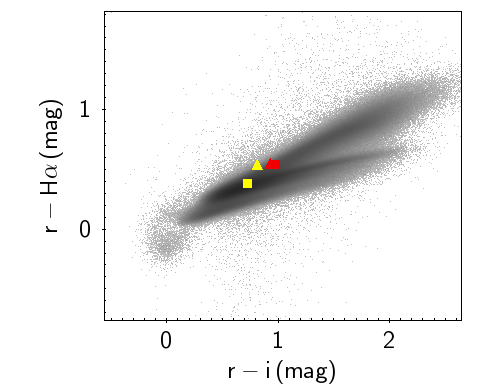}
    \caption{\normalsize Positions in the \gaia\ CAMD (left panel) and in the IPHAS TCD (right panel) of the objects in Fig.~\ref{fig:general_spectra}. The squares represent the objects with a significance (either $\sigma_{CAMD}$ and/or $\sigma_{POS}$) included between 3 and 5, while the triangles depict the sources with a higher significance. The colour-code is: yellow for the objects whose spectra show absorption in the \ha\ band, and red for the objects that show \ha\ emission.}
    \label{fig:general_spectra_CMD_CCD}
\end{figure*}

In Fig.~\ref{fig:general_spectra}, four spectra associated to our outliers are shown: two of these spectra belong to sources with an associated $\sigma$ included between 3 and 5, while the other two belong to objects with a higher significance. For each of these significance-based outliers sub-samples, one example of confirmed \ha-excess, and one example of clear absorption in the \ha\ band are shown. These four sources are located in the \gaia\ CAMD and in the IPHAS TCD in the left panel and in the right panel in Fig.~\ref{fig:general_spectra_CMD_CCD}, respectively.

\subsubsection{Spectral analysis of the cross-matches with Witham and IGAPS}

Out of the 98 Witham's outliers that are not identified by our selection (see Sec.~\ref{sec: comparison_witham}), 10 have an associated LAMOST DR5 spectrum. Four of these spectra show clear a \ha\ emission line. Overall, 533 spectra relative to the outliers in \cite{witham} are present in the LAMOST archive, and 481 of them (90.2\%) show a clear \ha\ emission line.

Regarding IGAPS $3\sigma$ outliers, 543 of them have an associated LAMOST DR5 spectrum, and 491 of these spectra (90.4\%) show \ha\ emission. Of the 843 IGAPS $3\sigma$ outliers that our algorithm does not identify as \ha\ emitting candidates (see Sec.~\ref{sec: comparison_IGAPS}), 21 have an associated LAMOST DR5 spectrum. The absolute majority of these spectra (18/21) shows a clear \ha\ emission line. On the other hand, out of the 3,417 3$\sigma$ positional-based outliers not included in IGAPS outliers list, 149 have an associated LAMOST spectrum. By visually inspecting these spectra, 57 of them (38.3\%) belong to clear \ha-excess sources. If constraining the subset to our 5$\sigma$ positional-outliers, 6 out of 9 available spectra show clear \ha\ emission.

\subsection{Cross-matches with faint-\textit{ROSAT}, bright-\textit{ROSAT} and CSC}

Accretion onto compact objects is often accompanied by X-rays emission. Table~\ref{tab:Xray_matches} provides the results of the cross-matches (with a radius of 15\,arcseconds) between our catalogue and three X-ray surveys: the \textit{ROSAT} All-Sky Survey Faint Source Catalogue (faint-\textit{ROSAT}, \citealp{faint_rosat}), the \textit{ROSAT} All-Sky Survey Bright Source Catalogue (bright-\textit{ROSAT}, \citealp{bright_rosat}) and the The Chandra Source Catalogue (CSC, \citealp{chandra_bis}). Among the 972 matches with faint-\textit{ROSAT}, 33 are identified as \ha-excess candidates by our algorithm. Almost all of these objects find a classification in SIMBAD (32/33): 30 out of 32 are identified as ``X-ray emitting sources\footnote{As the label ``Star", the ``X-ray emitting source" generic label in SIMBAD does not provide any further specification on the object being classified.}", and the remaining two as CVs. LAMOST spectra are available for three of these targets, and they all present a clear \ha\ emission line. Our algorithm assigns a significance (either CAMD-based and/or positional-based) higher than 5 to 12 of these 32 objects, including the two CVs.

10 $3\sigma$ outliers are included in the 69 matches with bright-\textit{ROSAT}, and all of them find a classification in SIMBAD. Three of them are classified as ``X-ray emitting source", three as CVs, two as Dwarf Novae, one WD, and one T-Tauri star. Three LAMOST spectra are available for this group of sources, as well; they all show the \ha\ line in emission; these spectra belong to the identified WD (of spectral type DA), the CV and the Dwarf Nova. Our algorithm associates a significance higher than 5 to eight of these 10 objects; only the WD and one of the source classified as ``X-ray emitting source" have a lower significance.

The cross-match between our catalogue and CSC yields 6,667 matches, 426 of which are identified as outliers by our algorithm. Out of these 426 objects, 342 are classified in SIMBAD. Among them, 264 are YSOs (or candidates) and T-Tauri stars, 29 are classified as ``Stars", 27 as Emission-line stars, and 15 as Orion Variables. Seven of these 342 objects find a LAMOST spectrum, and the \ha\ emission line is visible in all of them. 206 of these 342 targets have a significance higher than 5, and the vast majority of them (171/206) are classified in SIMBAD as YSOs.

\begin{figure*}
	\includegraphics[width=5\columnwidth/6, trim= 0mm 0mm 0mm 0mm]{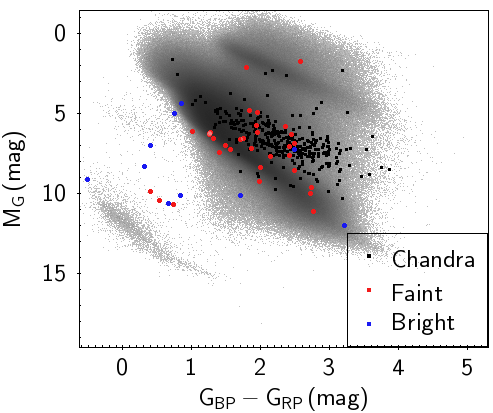}
	\includegraphics[width=5\columnwidth/6, trim= 0mm 0mm 0mm 0mm]{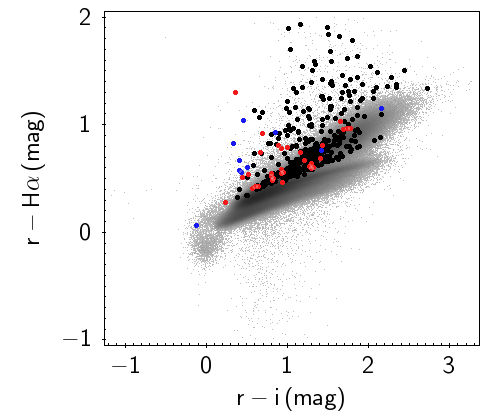}
    \caption{\normalsize The left panel presents the positions in the \gaia\ CAMD of our $r-$\ha\ $3\sigma$ outliers, that are included in the surveys bright-\textit{ROSAT} (\protect\citealp{bright_rosat}, the blue points), faint-\textit{ROSAT} (\protect\citealp{faint_rosat}, the red dots), and CSC (\protect\citealp{chandra_bis}, the black dots), for which a SIMBAD classification is available. The right panel shows the positions of the same objects in the IPHAS TCD.}
    \label{fig:rosat_simbad_positions}
\end{figure*}

In Fig.~\ref{fig:rosat_simbad_positions}, these three groups of 342, 32 and 10 sources are located in the \gaia\ CAMD (left panel) and in the IPHAS TCD (right panel). In agreement with their SIMBAD classification, most of the matches with CSC cluster around the area of the CAMD in which young accreting objects are expected to lie. Some of the sources in the two \textit{ROSAT} surveys are located between the MS track and the WD track, making them robust CV candidates. These are either already identified as such in SIMBAD, or are classified with the general label of ``X-Ray emitting sources".

\section{Enabled science cases}
\label{sec: follow-ups}
Two examples of possible science cases for our meta-catalogue are presented here. The first one consists of the identification of previously undetected accreting WD candidates, and it hinges directly on the cross-match between our outliers and the three X-Ray surveys presented in the previous section. In fact, accreting WDs usually occupy a well known region in the CAMD (between the MS and the WD tracks), and are associated to \ha\ and X-Ray emission. As an example, Lan 23 (\citealp{lan}, \citealp{lan2}) is a well known WD that is identified as an \ha\ outlier by our algorithm, its LAMOST spectrum shows an \ha\ emission line, and is found in the bright-\textit{ROSAT} catalogue.

\begin{figure}
	\includegraphics[width=6\columnwidth/7, trim= 0mm 0mm 0mm 0mm]{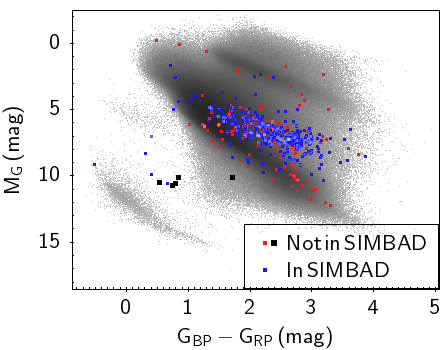}
    \caption{\normalsize Position in \gaia\ CAMD of our $3\sigma$ \ha\ outliers that show X-Ray emission. The blue dots represent the objects that are already classified in SIMBAD, while the red and black ones are unclassified (or simply classified as ``star" or ``X-Ray emitting source").}
    \label{fig:xray_outliers}
\end{figure}

Fig.~\ref{fig:xray_outliers} shows the position in \gaia\ CAMD of all our $3\sigma$ outliers that also present X-Ray emission. Among the X-Ray emitters that are not yet classified in SIMBAD, the black dots represent good examples of new robust accreting WD candidates. These objects are: \gaia\ DR2 414071753997318272, \gaia\ DR2 2060626872274773504, \gaia\ DR2 463350318963210624, \gaia\ DR2 2203244624288543744 and \gaia\ DR2 2203373473312011008.

\smallskip

\begin{figure}
	\includegraphics[width=6\columnwidth/7, trim= -4mm 0mm 3mm 0mm]{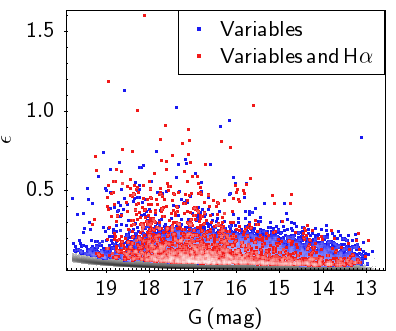}
	\includegraphics[width=6\columnwidth/7, trim= 0mm 0mm 0mm 0mm]{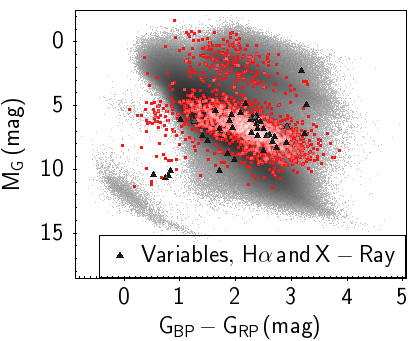}
    \caption{\normalsize Top panel: variable objects (the blue dots) and variable \ha\ outliers (the red dots) in our meta-catalogue. Bottom panel: position in the CAMD of our variable \ha-excess sources (the red dots). The black triangles represent the variable \ha\ outliers that also show X-Ray emission, and are not classified in SIMBAD.}
    \label{fig:variability}
\end{figure}

Another feature that characterises accreting WDs (as well as many other stellar populations) is variability. \cite{variability} developed a method that enables the calculation of a parameter ($\epsilon$) that quantifies the excess of Poissonian noise relative to the flux of a source. With the use of \gaia\ metrics, this parameter is given by:
\begin{equation}
    \epsilon=\sqrt{N} \cdot \frac{\delta f_G}{f_G}.
	\label{eq:variability}
\end{equation}
Here, f\textsubscript{G} is the mean G-band flux obtained with N observations, while $\delta$f\textsubscript{G} represents the corresponding dispersion. The sources in our catalogue are binned with respect to their G-band magnitude; the ones with an $\epsilon$ larger than five standard deviations above $\epsilon$\textsubscript{mean,i} (i.e. the average $\epsilon$ value for the i-th bin) are selected as variables. With this method, 22,199 variable sources are identified. According to our \ha-excess selection, 2,243 of them are also \ha-excess sources. The top panel of Fig.~\ref{fig:variability} shows the $\epsilon$ vs. G-band magnitude diagram: the blue dots represent the variable objects, whilst the red ones represent the variable \ha-excess sources. The bottom panel in the same figure shows the location of the variable \ha-excess sources in \gaia\ CAMD. While most of them cluster in the region associated to YSOs, many red dots are found in the region of the CAMD where B or A types stars lie, and between the MS and the WDs tracks.

A combination of X-Ray emission and variability makes our accreting WD candidates identification more robust. The bottom panel in Fig.~\ref{fig:variability} presents the position in the CAMD of the 41 variable objects that show X-Ray emission and are not yet classified in SIMBAD. The five black objects in Fig.~\ref{fig:xray_outliers} are included among them.




\section{Conclusions}
\label{sec: conclusion}

In this study, a new method for selecting \ha-excess candidates from a vast photometric survey is presented. Our analysis is performed on the \gIPHAS, produced by \cite{Scaringi}. It comprises targets included in the $|b|\leq 5$\textdegree\ and 29\textdegree $\leq l \leq$ 215\textdegree\ ranges, within a radius of $\sim1.5$\,kpc. \gaia\ photometric measurements and parallaxes play a key role in the development of our selection: by locating the sources in the \gaia\ CAMD, it is possible to associate them to a stellar population. In order to minimise the effects due to stellar population mixing, the targets are partitioned in the \gaia\ CAMD; to mitigate the effects of extintion, they are further (and independently) partitioned in the Galactic coordinates space. For each partition, the main locus in the IPHAS TCD, and subsequently the $r-$\ha\ outliers, are found by applying the iterative Chauvenet's criterion twice. The \ha-excess candidates are thus defined as the sources that satisfy the criterion in Eq.~\ref{eq:selection}.

This process leads to the identification a new set of \ha\-excess candidates in the Northern Galactic plane. In fact, the partition of the sources in two different parameter spaces enables the identification of \ha\ line candidates that would be otherwise hidden among different stellar populations. More specifically, 28,496 \ha-excess candidates (0.4\% of the total dataset) are identified, with either $\sigma_{CAMD}\geq 3$ and/or $\sigma_{POS}\geq 3$. However, a $5\sigma$ cut is suggested, as it constitutes a solid agreement between completeness and conservativeness. By applying this latter cut, 6,774 objects (23.8\% of the $\sim28,500$ $3\sigma$ outliers) are identified as \ha-excess sources.
The visual inspection of the available LAMOST DR5 spectra of our $3\sigma$ outliers shows that 48.9\% of them exhibit a clear \ha\ emission line. This purity fraction does not improve significantly if constraining the outliers to the $5\sigma$ subset: 49.7\% of them are confirmed \ha\ emitters by the available LAMOST spectra. These apparently low percentages are explained by the fact that our algorithm identifies \ha-\textit{excess} sources, rather than \ha\ \textit{emitters}. This is also consistent with the spectral confirmation rate being systematically higher for our positional-outliers than for our CAMD-outliers. However, by retaining only the outliers that are at least half magnitude fainter than IPHAS saturation limits, 67.7\% of our $3\sigma$ outliers - and 81.9\% of our $5\sigma$ outliers - are spectroscopically confirmed as reliable \ha-excess sources. This latter selection cuts are therefore suggested.

Our $3\sigma$ selection identifies between 3\% and 5\% of the \ha\ emitters in the Northern Galactic plane. Despite this constitutes an improvement with respect to previous similar studies, it also suggests that our knowledge of the \ha\ emitters in the Galaxy is still far from being complete.

The results of our analysis are presented in our \textit{meta-catalogue} of the \gaia/IPHAS \ha-excess sources. This includes all the 7,474,835 objects in the master-catalogue, and for each of them, the following specifications are provided: the \gaia\ DR2 SourceID, the equatorial coordinates, the distance, IPHAS DR2 and \gaia\ DR2 photometric measurements, as well as the two $flagCAMD$ and $flagPOS$ labels, that specify the confidence level that the source is an \ha-excess candidate.  Moreover, a full-version of our catalogue is available, in which the whole set of metrics obtained during the \ha-excess selection process is added.

A cross-match with SIMBAD \citep{simbad} shows that 6.4\% of our $3\sigma$ outliers have been previously identified. However, if followed up spectroscopically, our list of outliers can be used to enhance the census of identified \ha\ emitting point-like sources, such as CVs or SySts. This constitutes a profitable starting point to address, for instance, the problem of the difference between observed and predicted CVs spatial density in the Galactic plane (\citealt{de_kool}, \citealt{kolb}). Although \cite{belloni2020} seem to have found a promising way to overcome this impasse, their conclusions are still to be confirmed \citep{pala}. Moreover, the identification of new \ha\ emitting sources can foster population studies, which by definition need a vast amount of objects to be performed. In addition, newly classified sources can provide further pieces of the puzzle for a better understanding about the possible evolutionary models of the stellar population they belong to.

With the arrival of \gaia\ early Data Release 3 (\gaia\ eDR3), our intention is to apply our analysis on the list of objects resulting from the cross-match between \gaia\ eDR3 and IGAPS. This will provide more up to date results, compared to the ones in our current meta-catalogue.

\section{Data availability}
Both the light and the full versions of our meta-catalogue can be found in VizieR as ``The \gaia/IPHAS catalogue of \ha-excess sources".

\begin{landscape}

\begin{table}\centering
 \begin{tabular}{|c|c|c|c|c|c|c|c|c|c|c|c|c|c|c|}
  \hline
  \shortstack{\textbf{SourceID} \\ (\gaia\ $DR2$) } & \shortstack{\textbf{RA}\\ (\textdegree)} & \shortstack{\textbf{Dec}\\ (\textdegree)} & \shortstack{\textbf{Distance}\\ (kpc)} & \shortstack{\textbf{$r$}\\(mag)} & \shortstack{\textbf{e\_$r$}\\ (mag)} & \shortstack{\textbf{$i$}\\ (mag)} & \shortstack{\textbf{e\_$i$}\\ (mag)} & \shortstack{\textbf{\ha}\\(mag)}  & \shortstack{\textbf{e\_\ha}\\ (mag)} & \shortstack{\textbf{G\textsubscript{BP}}\\ (mag)} & \shortstack{\textbf{G\textsubscript{RP}}\\ (mag)} & \shortstack{\textbf{M\textsubscript{G}}\\ (mag)} & \shortstack{\textbf{flag}\\ \textbf{CAMD}} & \shortstack{\textbf{flag}\\ \textbf{POS}}  \\
  \hline\hline
  429950213735495552 & 0.01155 & 62.53021 & 4.91191 & 13.217 & 0.001 & 12.919 & 0.002 & 12.963 & 0.001 & 13.418 & 12.938 & -0.196 & 0 & 1\\
  \hline
  430049096757310848 & 0.04097 & 62.80630 & 0.24588 & 17.815 & 0.017 & 17.868 & 0.026 & 17.663 & 0.018 & 17.612 & 17.808 & 10.816 & 0 &1\\
  \hline
  432344808313438336 & 0.04268 & 66.34893 & 0.72553 & 18.881 & 0.016 & 17.163 & 0.014 & 17.851 & 0.019 & 20.173 & 16.946 & 9.068 & 1 &1\\
  \hline
  432167168466312448 & 0.05005 & 65.17739 & 0.80185 & 18.114 & 0.009 & 16.901 & 0.012 & 17.378 & 0.013 & 19.016 & 16.783 & 8.374 & 0 &1\\
  \hline
  429950179375766144 & 0.06964 & 62.53050 & 0.44258 & 13.017 & 0.001 & 12.419 & 0.001 & 12.593 & 0.001 & 13.477 & 12.461 & 4.819 & 2 &1\\
  \hline
  429952687636601728 & 0.09328 & 62.64748 & 0.32689 & 13.639 & 0.001 & 13.100 & 0.002 & 13.276 & 0.002 & 14.408 & 13.114 & 6.252 & 0 &1\\
 \hline
 429521606051123712 & 0.11978 & 61.69365 & 2.61161 & 16.222 & 0.003 & 15.723 & 0.006 & 15.738 & 0.006 & 16.619 & 15.392 & 4.000 & 1 &2\\
 \hline
 432168650239229696 & 0.12183 & 65.20698 & 0.50034 & 13.080 & 0.001 & 12.362 & 0.001 & 12.617 & 0.001 & 13.615 & 12.402 & 4.591 & 2 &0\\
 \hline
 429915231214176768 & 0.12447 & 62.17243 & 1.18993 & 17.926 & 0.010 & 16.656 & 0.011 & 17.262 & 0.015 & 18.942 & 16.643 & 7.368 & 1 &0\\
 \hline
 430048723107383808 & 0.12939 & 62.77123 & 1.20174 & 15.160 & 0.004 & 14.507 & 0.003 & 14.679 & 0.003 & 15.498 & 14.192 & 4.520 & 2 &1\\
 \hline
  
 \shortstack{ .\\.\\.} & \shortstack{ .\\.\\.} & \shortstack{ .\\.\\.} & \shortstack{ .\\.\\.} & \shortstack{ .\\.\\.} & \shortstack{ .\\.\\.} & \shortstack{ .\\.\\.} & \shortstack{ .\\.\\.} & \shortstack{ .\\.\\.} & \shortstack{ .\\.\\.} & \shortstack{ .\\.\\.} & \shortstack{ .\\.\\.} & \shortstack{ .\\.\\.} & \shortstack{ .\\.\\.} & \shortstack{ .\\.\\.} \\
 \end{tabular}
 \caption{\normalsize The table shows the first ten rows of our \textit{meta-catalogue}. For each source, the following entries are provided: the \gaia\ DR2 SourceID; the \gaia\ DR2 barycentric equatorial coordinates at epoch 2015.5; the distance (calculated in \protect\citealp{Scaringi}); the IPHAS DR2 photometric measurements with corresponding errors; the \gaia\ DR2 photometric measurements; the two labels \textit{flagCAMD} and \textit{flagPOS}. These two latter parameters express how likely each source is to be an outlier, within its corresponding $r-$\ha\ distribution (either in the CAMD-partitions and/or in the positional-partitions).}
 \label{tab:light_catalogue}
\end{table}

\hspace{1cm}

\begin{table}\centering
 \begin{tabular}{|c||c|c|c|c|c|c|}
  \hline
   
  & \shortstack{\textbf{Total} \\\textbf{}}& \shortstack{\textbf{$flagCAMD =0$} \\{\textbf{$flagPOS =0$}} } &\shortstack{\textbf{$flagCAMD =1$} \\\textbf{}} &\shortstack{\textbf{$flagCAMD =2$} \\\textbf{}} & \shortstack{\textbf{$flagPOS =1$} \\\textbf{}} & \shortstack{\textbf{$flagPOS =2$} \\\textbf{}}  \\
  \hline\hline
  \shortstack{\textbf{faint} \\{\textbf{\textit{ROSAT}}}} & 972 & 939  (96.6\%) & 26  (2.7\%)& 10 (1.0\%)& 17 (1.7\%)& 8 (0.8\%)\\
  \hline
  \shortstack{\textbf{bright} \\{\textbf{\textit{ROSAT}}}} & 69 & 59 (85.5\%)& 8 (11.6\%) & 8 (11.6\%)& 9 (13.0\%)& 7 (10.1\%)\\
  \hline
  \shortstack{\textbf{ } \\ {\textbf{CSC}}} & 6,667 & 6,241 (93.6\%) & 177 (2.7\%) & 218 (3.3\%) & 71 (1.1\%) & 72 (1.1\%) \\
  \hline
 
 \end{tabular}
 \caption{\normalsize The table shows the results of different cross-matches between our meta-catalogue and the three X-rays surveys faint-\textit{ROSAT} \protect\citep{faint_rosat}, bright-\textit{ROSAT} \protect\citep{bright_rosat} and CSC \protect\citep{chandra_bis}. The objects in our dataset are grouped before the cross-matches, with reference to the corresponding flagCAMD and flagPOS specifications. These entries refer to the significances (either CAMD-based or positional-based) associated to each object in the catalogue.}
 \label{tab:Xray_matches}
\end{table}
\end{landscape}

\section*{Acknowledgements}
The cross-matches, as well as some of the figures used for this paper were produced with the use of the astronomy-oriented software ``TOPCAT" \citep{topcat}. MM acknowledges the support by the Spanish Ministry of Science, Innovation and University (MICIU/FEDER, UE) through grant RTI2018-095076-B-C21, and the Institute of Cosmos Sciences University of Barcelona (ICCUB, Unidad de Excelencia ``Mar\'{\i}a de Maeztu") through grant CEX2019-000918-M.





\bibliographystyle{mnras}
\bibliography{refs.bib}

\begin{thebibliography}{}
\makeatletter
\relax
\def\mn@urlcharsother{\let\do\@makeother \do\$\do\&\do\#\do\^\do\_\do\%\do\~}
\def\mn@doi{\begingroup\mn@urlcharsother \@ifnextchar [ {\mn@doi@}
  {\mn@doi@[]}}
\def\mn@doi@[#1]#2{\def\@tempa{#1}\ifx\@tempa\@empty \href
  {http://dx.doi.org/#2} {doi:#2}\else \href {http://dx.doi.org/#2} {#1}\fi
  \endgroup}
\def\mn@eprint#1#2{\mn@eprint@#1:#2::\@nil}
\def\mn@eprint@arXiv#1{\href {http://arxiv.org/abs/#1} {{\tt arXiv:#1}}}
\def\mn@eprint@dblp#1{\href {http://dblp.uni-trier.de/rec/bibtex/#1.xml}
  {dblp:#1}}
\def\mn@eprint@#1:#2:#3:#4\@nil{\def\@tempa {#1}\def\@tempb {#2}\def\@tempc
  {#3}\ifx \@tempc \@empty \let \@tempc \@tempb \let \@tempb \@tempa \fi \ifx
  \@tempb \@empty \def\@tempb {arXiv}\fi \@ifundefined
  {mn@eprint@\@tempb}{\@tempb:\@tempc}{\expandafter \expandafter \csname
  mn@eprint@\@tempb\endcsname \expandafter{\@tempc}}}

\bibitem[\protect\citeauthoryear{{Abrahams}, {Bloom}, {Mowlavi}, {Szkody},
  {Rix}, {Ventura}, {Brink}  \& {Filippenko}}{{Abrahams}
  et~al.}{2020}]{variability}
{Abrahams} E.~S.,  {Bloom} J.~S.,  {Mowlavi} N.,  {Szkody} P.,  {Rix} H.-W.,
  {Ventura} J.-P.,  {Brink} T.~G.,   {Filippenko} A.~V.,  2020, arXiv e-prints,
  \href {https://ui.adsabs.harvard.edu/abs/2020arXiv201112253A} {p.
  arXiv:2011.12253}

\bibitem[\protect\citeauthoryear{{Astraatmadja} \&
  {Bailer-Jones}}{{Astraatmadja} \& {Bailer-Jones}}{2016}]{dist}
{Astraatmadja} T.~L.,  {Bailer-Jones} C. A.~L.,  2016, \mn@doi [\apj]
  {10.3847/1538-4357/833/1/119}, \href
  {https://ui.adsabs.harvard.edu/abs/2016ApJ...833..119A} {833, 119}

\bibitem[\protect\citeauthoryear{{Barentsen}, {Vink}, {Drew}  \&
  {Sale}}{{Barentsen} et~al.}{2013}]{NGC2264}
{Barentsen} G.,  {Vink} J.~S.,  {Drew} J.~E.,   {Sale} S.~E.,  2013, \mn@doi
  [\mnras] {10.1093/mnras/sts462}, \href
  {https://ui.adsabs.harvard.edu/abs/2013MNRAS.429.1981B} {429, 1981}

\bibitem[\protect\citeauthoryear{{Barentsen} et~al.,}{{Barentsen}
  et~al.}{2014}]{barentsen}
{Barentsen} G.,  et~al., 2014, \mn@doi [\mnras] {10.1093/mnras/stu1651}, \href
  {https://ui.adsabs.harvard.edu/abs/2014MNRAS.444.3230B} {444, 3230}

\bibitem[\protect\citeauthoryear{{Belloni}, {Schreiber}, {Pala},
  {G{\"a}nsicke}, {Zorotovic}  \& {Rodrigues}}{{Belloni}
  et~al.}{2020}]{belloni2020}
{Belloni} D.,  {Schreiber} M.~R.,  {Pala} A.~F.,  {G{\"a}nsicke} B.~T.,
  {Zorotovic} M.,   {Rodrigues} C.~V.,  2020, \mn@doi [\mnras]
  {10.1093/mnras/stz3413}, \href
  {https://ui.adsabs.harvard.edu/abs/2020MNRAS.491.5717B} {491, 5717}

\bibitem[\protect\citeauthoryear{{Bressan}, {Marigo}, {Girardi}, {Salasnich},
  {Dal Cero}, {Rubele}  \& {Nanni}}{{Bressan} et~al.}{2012}]{Bressan}
{Bressan} A.,  {Marigo} P.,  {Girardi} L.,  {Salasnich} B.,  {Dal Cero} C.,
  {Rubele} S.,   {Nanni} A.,  2012, \mn@doi [\mnras]
  {10.1111/j.1365-2966.2012.21948.x}, \href
  {http://adsabs.harvard.edu/abs/2012MNRAS.427..127B} {427, 127}

\bibitem[\protect\citeauthoryear{{Carrasco}, {Catal{\'a}n}, {Jordi},
  {Tremblay}, {Napiwotzki}, {Luri}, {Robin}  \& {Kowalski}}{{Carrasco}
  et~al.}{2014}]{Carrasco}
{Carrasco} J.~M.,  {Catal{\'a}n} S.,  {Jordi} C.,  {Tremblay} P.-E.,
  {Napiwotzki} R.,  {Luri} X.,  {Robin} A.~C.,   {Kowalski} P.~M.,  2014,
  \mn@doi [\aap] {10.1051/0004-6361/201220596}, \href
  {http://adsabs.harvard.edu/abs/2014A%26A...565A..11C} {565, A11}

\bibitem[\protect\citeauthoryear{{Cutri} et~al.,}{{Cutri} et~al.}{2003}]{lan2}
{Cutri} R.~M.,  et~al., 2003, VizieR Online Data Catalog, \href
  {https://ui.adsabs.harvard.edu/abs/2003yCat.2246....0C} {p. II/246}

\bibitem[\protect\citeauthoryear{{Davies}, {Elliott}  \& {Meaburn}}{{Davies}
  et~al.}{1976}]{davies}
{Davies} R.~D.,  {Elliott} K.~H.,   {Meaburn} J.,  1976, \memras, \href
  {https://ui.adsabs.harvard.edu/abs/1976MmRAS..81...89D} {81, 89}

\bibitem[\protect\citeauthoryear{{Dennison}, {Simonetti}  \&
  {Topasna}}{{Dennison} et~al.}{1999}]{dennison}
{Dennison} B.,  {Simonetti} J.~H.,   {Topasna} G.~A.,  1999, in American
  Astronomical Society Meeting Abstracts. p. 53.09

\bibitem[\protect\citeauthoryear{{Dias}, {Monteiro}, {Caetano}, {L{\'e}pine},
  {Assafin}  \& {Oliveira}}{{Dias} et~al.}{2014}]{NGC2264bis}
{Dias} W.~S.,  {Monteiro} H.,  {Caetano} T.~C.,  {L{\'e}pine} J.~R.~D.,
  {Assafin} M.,   {Oliveira} A.~F.,  2014, \mn@doi [\aap]
  {10.1051/0004-6361/201323226}, \href
  {https://ui.adsabs.harvard.edu/abs/2014A&A...564A..79D} {564, A79}

\bibitem[\protect\citeauthoryear{{Drew} et~al.,}{{Drew} et~al.}{2005}]{drew}
{Drew} J.~E.,  et~al., 2005, \mn@doi [\mnras]
  {10.1111/j.1365-2966.2005.09330.x}, \href
  {http://adsabs.harvard.edu/abs/2005MNRAS.362..753D} {362, 753}

\bibitem[\protect\citeauthoryear{{Drew} et~al.,}{{Drew} et~al.}{2014}]{vphas}
{Drew} J.~E.,  et~al., 2014, \mn@doi [\mnras] {10.1093/mnras/stu394}, \href
  {https://ui.adsabs.harvard.edu/abs/2014MNRAS.440.2036D} {440, 2036}

\bibitem[\protect\citeauthoryear{{Evans} et~al.,}{{Evans}
  et~al.}{2010}]{chandra_bis}
{Evans} I.~N.,  et~al., 2010, \mn@doi [\apjs] {10.1088/0067-0049/189/1/37},
  \href {https://ui.adsabs.harvard.edu/abs/2010ApJS..189...37E} {189, 37}

\bibitem[\protect\citeauthoryear{{Farnhill}, {Drew}, {Barentsen}  \&
  {Gonz{\'a}lez-Solares}}{{Farnhill} et~al.}{2016}]{farnhill}
{Farnhill} H.~J.,  {Drew} J.~E.,  {Barentsen} G.,   {Gonz{\'a}lez-Solares}
  E.~A.,  2016, \mn@doi [\mnras] {10.1093/mnras/stv2994}, \href
  {https://ui.adsabs.harvard.edu/abs/2016MNRAS.457..642F} {457, 642}

\bibitem[\protect\citeauthoryear{{Gaustad}, {McCullough}, {Rosing}  \& {Van
  Buren}}{{Gaustad} et~al.}{2001}]{gaustad}
{Gaustad} J.~E.,  {McCullough} P.~R.,  {Rosing} W.,   {Van Buren} D.,  2001,
  \mn@doi [\pasp] {10.1086/323969}, \href
  {https://ui.adsabs.harvard.edu/abs/2001PASP..113.1326G} {113, 1326}

\bibitem[\protect\citeauthoryear{{Glazebrook}, {Peacock}, {Collins}  \&
  {Miller}}{{Glazebrook} et~al.}{1994}]{glazebrook}
{Glazebrook} K.,  {Peacock} J.~A.,  {Collins} C.~A.,   {Miller} L.,  1994,
  \mn@doi [\mnras] {10.1093/mnras/266.1.65}, \href
  {https://ui.adsabs.harvard.edu/abs/1994MNRAS.266...65G} {266, 65}

\bibitem[\protect\citeauthoryear{{Groot} et~al.,}{{Groot} et~al.}{2009}]{uvex}
{Groot} P.~J.,  et~al., 2009, \mn@doi [\mnras]
  {10.1111/j.1365-2966.2009.15273.x}, \href
  {https://ui.adsabs.harvard.edu/abs/2009MNRAS.399..323G} {399, 323}

\bibitem[\protect\citeauthoryear{{Irwin} \& {Lewis}}{{Irwin} \&
  {Lewis}}{2001}]{irwin}
{Irwin} M.,  {Lewis} J.,  2001, \mn@doi [\nar] {10.1016/S1387-6473(00)00138-X},
  \href {https://ui.adsabs.harvard.edu/abs/2001NewAR..45..105I} {45, 105}

\bibitem[\protect\citeauthoryear{{Kharchenko}, {Piskunov}, {Schilbach},
  {R{\"o}ser}  \& {Scholz}}{{Kharchenko} et~al.}{2013}]{IC1396}
{Kharchenko} N.~V.,  {Piskunov} A.~E.,  {Schilbach} E.,  {R{\"o}ser} S.,
  {Scholz} R.~D.,  2013, \mn@doi [\aap] {10.1051/0004-6361/201322302}, \href
  {https://ui.adsabs.harvard.edu/abs/2013A&A...558A..53K} {558, A53}

\bibitem[\protect\citeauthoryear{{Kohoutek} \& {Wehmeyer}}{{Kohoutek} \&
  {Wehmeyer}}{1999}]{kw99}
{Kohoutek} L.,  {Wehmeyer} R.,  1999, \mn@doi [\aaps] {10.1051/aas:1999101},
  \href {https://ui.adsabs.harvard.edu/abs/1999A&AS..134..255K} {134, 255}

\bibitem[\protect\citeauthoryear{{Kolb}}{{Kolb}}{1993}]{kolb}
{Kolb} U.,  1993, \aap, \href
  {https://ui.adsabs.harvard.edu/abs/1993A&A...271..149K} {271, 149}

\bibitem[\protect\citeauthoryear{{Kuhn}, {Hillenbrand}, {Sills}, {Feigelson}
  \& {Getman}}{{Kuhn} et~al.}{2019}]{NGC2264_coords}
{Kuhn} M.~A.,  {Hillenbrand} L.~A.,  {Sills} A.,  {Feigelson} E.~D.,   {Getman}
  K.~V.,  2019, \mn@doi [\apj] {10.3847/1538-4357/aaef8c}, \href
  {https://ui.adsabs.harvard.edu/abs/2019ApJ...870...32K} {870, 32}

\bibitem[\protect\citeauthoryear{{Lindegren} et~al.,}{{Lindegren}
  et~al.}{2018}]{lindgren}
{Lindegren} L.,  et~al., 2018, \mn@doi [\aap] {10.1051/0004-6361/201832727},
  \href {https://ui.adsabs.harvard.edu/abs/2018A&A...616A...2L} {616, A2}

\bibitem[\protect\citeauthoryear{{Luo}, {Zhao}, {Zhao}  \& {et al.}}{{Luo}
  et~al.}{2019}]{2019lamost}
{Luo} A.~L.,  {Zhao} Y.~H.,  {Zhao} G.,   {et al.} 2019, VizieR Online Data
  Catalog, \href {https://ui.adsabs.harvard.edu/abs/2019yCat.5164....0L} {p.
  V/164}

\bibitem[\protect\citeauthoryear{{Magnier} et~al.,}{{Magnier}
  et~al.}{2013}]{magnier}
{Magnier} E.~A.,  et~al., 2013, \mn@doi [\apjs] {10.1088/0067-0049/205/2/20},
  \href {https://ui.adsabs.harvard.edu/abs/2013ApJS..205...20M} {205, 20}

\bibitem[\protect\citeauthoryear{{Mohr-Smith} et~al.,}{{Mohr-Smith}
  et~al.}{2017}]{mohr}
{Mohr-Smith} M.,  et~al., 2017, \mn@doi [\mnras] {10.1093/mnras/stw2751}, \href
  {https://ui.adsabs.harvard.edu/abs/2017MNRAS.465.1807M} {465, 1807}

\bibitem[\protect\citeauthoryear{{Mongui{\'o}} et~al.,}{{Mongui{\'o}}
  et~al.}{2020}]{monguio}
{Mongui{\'o}} M.,  et~al., 2020, \mn@doi [\aap] {10.1051/0004-6361/201937333},
  \href {https://ui.adsabs.harvard.edu/abs/2020A&A...638A..18M} {638, A18}

\bibitem[\protect\citeauthoryear{{Pala} et~al.,}{{Pala} et~al.}{2020}]{pala}
{Pala} A.~F.,  et~al., 2020, \mn@doi [\mnras] {10.1093/mnras/staa764}, \href
  {https://ui.adsabs.harvard.edu/abs/2020MNRAS.494.3799P} {494, 3799}

\bibitem[\protect\citeauthoryear{{Parker} et~al.,}{{Parker} et~al.}{2005}]{shs}
{Parker} Q.~A.,  et~al., 2005, \mn@doi [\mnras]
  {10.1111/j.1365-2966.2005.09350.x}, \href
  {https://ui.adsabs.harvard.edu/abs/2005MNRAS.362..689P} {362, 689}

\bibitem[\protect\citeauthoryear{{Raddi} et~al.,}{{Raddi} et~al.}{2013}]{raddi}
{Raddi} R.,  et~al., 2013, \mn@doi [\mnras] {10.1093/mnras/stt038}, \href
  {https://ui.adsabs.harvard.edu/abs/2013MNRAS.430.2169R} {430, 2169}

\bibitem[\protect\citeauthoryear{{Sale} et~al.,}{{Sale} et~al.}{2014}]{sale}
{Sale} S.~E.,  et~al., 2014, \mn@doi [\mnras] {10.1093/mnras/stu1090}, \href
  {https://ui.adsabs.harvard.edu/abs/2014MNRAS.443.2907S} {443, 2907}

\bibitem[\protect\citeauthoryear{{Scaringi} et~al.,}{{Scaringi}
  et~al.}{2018}]{Scaringi}
{Scaringi} S.,  et~al., 2018, \mn@doi [\mnras] {10.1093/mnras/sty2498}, \href
  {http://adsabs.harvard.edu/abs/2018MNRAS.481.3357S} {481, 3357}

\bibitem[\protect\citeauthoryear{{Taylor}}{{Taylor}}{2005}]{topcat}
{Taylor} M.~B.,  2005, in {Shopbell} P.,  {Britton} M.,   {Ebert} R.,  eds,
  Astronomical Society of the Pacific Conference Series Vol. 347, Astronomical
  Data Analysis Software and Systems XIV. p.~29

\bibitem[\protect\citeauthoryear{{Voges} et~al.,}{{Voges}
  et~al.}{1999}]{bright_rosat}
{Voges} W.,  et~al., 1999, \aap, \href
  {https://ui.adsabs.harvard.edu/abs/1999A&A...349..389V} {349, 389}

\bibitem[\protect\citeauthoryear{{Voges} et~al.,}{{Voges}
  et~al.}{2000}]{faint_rosat}
{Voges} W.,  et~al., 2000, VizieR Online Data Catalog, \href
  {https://ui.adsabs.harvard.edu/abs/2000yCat.9029....0V} {p. IX/29}

\bibitem[\protect\citeauthoryear{{Wenger} et~al.,}{{Wenger}
  et~al.}{2000}]{simbad}
{Wenger} M.,  et~al., 2000, \mn@doi [\aaps] {10.1051/aas:2000332}, \href
  {https://ui.adsabs.harvard.edu/abs/2000A&AS..143....9W} {143, 9}

\bibitem[\protect\citeauthoryear{{Witham}, {Knigge}, {Drew}, {Greimel},
  {Steeghs}, {G{\"a}nsicke}, {Groot}  \& {Mampaso}}{{Witham}
  et~al.}{2008}]{witham}
{Witham} A.~R.,  {Knigge} C.,  {Drew} J.~E.,  {Greimel} R.,  {Steeghs} D.,
  {G{\"a}nsicke} B.~T.,  {Groot} P.~J.,   {Mampaso} A.,  2008, \mn@doi [\mnras]
  {10.1111/j.1365-2966.2007.12774.x}, \href
  {http://adsabs.harvard.edu/abs/2008MNRAS.384.1277W} {384, 1277}

\bibitem[\protect\citeauthoryear{{Wramdemark}}{{Wramdemark}}{1981}]{lan}
{Wramdemark} S.,  1981, \aaps, \href
  {https://ui.adsabs.harvard.edu/abs/1981A&AS...43..103W} {43, 103}

\bibitem[\protect\citeauthoryear{{de Kool}}{{de Kool}}{1992}]{de_kool}
{de Kool} M.,  1992, \aap, \href
  {https://ui.adsabs.harvard.edu/abs/1992A&A...261..188D} {261, 188}

\makeatother
\end{thebibliography}



\appendix




\bsp	
\label{lastpage}
\end{document}